\documentclass[twocolumn]{aastex62}
\usepackage{graphicx}	
\usepackage{amsmath}	
\usepackage{amssymb}	
\usepackage{wasysym}
\usepackage{savesym}
\savesymbol{tablenum}
\usepackage[separate-uncertainty=true,multi-part-units=single,detect-weight=true, detect-family=true]{siunitx}
\restoresymbol{SIX}{tablenum}
\DeclareSIUnit\year{yr}
\DeclareSIUnit\erg{erg}
\urldef\moldataurl\url{http://home.strw.leidenuniv.nl/~moldata/}
\received{January 1, 2018}
\revised{January 7, 2018}
\accepted{\today}
\submitjournal{ApJ}
\shorttitle{Sulfur in L1157-B1}
\shortauthors{Holdship et al.}

\begin{document}

\title{Sulfur Chemistry in L1157-B1}

\correspondingauthor{Izaskun Jimenez-Serra}
\email{ijimenez@cab.inta-csic.es}

\author[0000-0003-4025-1552]{Jonathan Holdship}
\affiliation{Department of Physics and Astronomy, University College London, Gower Street, London WC1E 6BT}

\author{Izaskun Jimenez-Serra}
\affiliation{Centro de Astrobiología (CSIC, INTA), Ctra. de Ajalvir, km. 4, Torrejón de Ardoz, E-28850 Madrid, Spain}

\author[0000-0001-8504-8844]{Serena Viti}
\affiliation{Department of Physics and Astronomy, University College London, Gower Street, London WC1E 6BT}

\author{Claudio Codella}
\affiliation{INAF, Osservatorio Astrofisico di Arcetri, Largo E. Fermi 5, 50125 Firenze, Italy}
\affiliation{Univ. Grenoble Alpes, Institut de Planétologie et d'Astrophysique
de Grenoble (IPAG), 38401 Grenoble, France}

\author{Milena Benedettini}
\affiliation{INAF, Istituto di Astrofisica e Planetologia Spaziali, via Fosso del Cavaliere 100, I-00133 Roma, Italy}

\author{Francesco Fontani}
\affiliation{INAF, Osservatorio Astrofisico di Arcetri, Largo E. Fermi 5, 50125 Firenze, Italy}

\author[0000-0003-2733-5372]{Linda Podio}
\affiliation{INAF, Osservatorio Astrofisico di Arcetri, Largo E. Fermi 5, 50125 Firenze, Italy}

\author{Mario Tafalla}
\affiliation{IGN, Observatorio Astron\'omico Nacional, Calle Alfonso XII, E-28014 Madrid, Spain}

\author{Rafael Bachiller}
\affiliation{IGN, Observatorio Astron\'omico Nacional, Calle Alfonso XII, E-28014 Madrid, Spain}

\author{Cecilia Ceccarelli}
\affiliation{Univ. Grenoble Alpes, Institut de Plan\'etologie et d'Astrophysique de Grenoble (IPAG), 38401 Grenoble, France}
\affiliation{CNRS, Institut de Plan\'etologie et d'Astrophysique de Grenoble (IPAG), 38401 Grenoble, France}

\begin{abstract}
The main carrier of sulfur in dense clouds, where it is depleted from the gas phase, remains a mystery. Shock waves in young molecular outflows disrupt the ice mantles and allow us to directly probe the material that is ejected into the gas phase. A comprehensive study of sulfur-bearing species towards L1157-B1, a shocked region along a protostellar outflow, has been carried out as part of the IRAM-30m large program ASAI. The dataset contains over 100 lines of CCS, H$_2$CS, OCS, SO, SO$_2$ and isotopologues. The results of these observations are presented, complementing previous studies of sulfur-bearing species in the region. The column densities and fractional abundances of these species are measured and together these species account for 10\% of the cosmic sulfur abundance in the region. The gas properties derived from the observations are also presented, demonstrating that sulfur bearing species trace a wide range of different gas conditions in the region.
\end{abstract}

\section{Introduction}
Jets are ejected from low mass protostars and collide with the surrounding gas at speeds orders of magnitude higher than the typical sound speed in a molecular cloud. The shocks formed from these collisions sputter the ice mantles of dust grains, ejecting molecules into the gas phase and greatly increasing the chemical complexity of the gas \citep[eg.][]{Draine1983,Codella2013}. Therefore, these outflows represent fantastic laboratories to study chemistry in shocks as well as on the grain surfaces.\par
L1157-mm is a low mass, class 0 protostar at a distance of 250 pc \citep{looney2007}. L1157-mm drives a precessing jet \citep{Gueth1996,Gueth1998,Podio2016}, which in turn accelerates an extended outflow which was found to contain many bow-shocks \citep{bachiller2001}, the brightest of which is L1157-B1 in the blue shifted lobe. It has a dynamical age of $\sim$1000 yr \citep{Podio2016} and is well studied. It has a rich chemistry, including species thought to be formed on the ice mantles such as CH$_3$OH \citep{Bachiller1997} making it a superb object in which to study chemistry in shocks.\par
L1157-B1 is a bow-shock with clear, clumpy substructure \citep{Benedettini2007}, in which different chemical species trace only the clumps or the region as a whole. L1157-B1 and, in particular, the clumps which compose it are rich in sulfur-bearing species. In the interferometry work of \citet{Benedettini2007}, B1 was shown to be well defined in OCS, $^{34}$SO and CS emission. Further, SO$^+$ and SiS were discovered for the first time in a shock in L1157-B1 \citep{Podio2014,Podio2017}, SO$^+$ being part of a search for molecular ions in which HCS$^+$ was also detected. Overall, L1157-B1 provides a rich source of data on sulfur-bearing species with which to advance our understanding of sulfur chemistry, particularly of the form of sulfur that has been depleted from the gas phase.\par
Sulfur is a reactive element with a poorly understood surface chemistry. Many species hydrogenate efficiently on the grains and therefore it is often assumed H$_2$S is a major carrier of sulfur on the grains. Indeed, recent modelling work has shown sulfur abundances in the TMC-1 cloud are best described when HS and H$_2$S on the grains are the main carrier of sulfur \citep{vidal2017}. However, upper limits have also been placed on the H$_2$S abundance in ices around high mass young stellar objects and it was found that H$_2$S had a solid phase abundance \textless 0.7\% of the solid water abundance \citep{Jimenez-escobar2011}. This would account for less than $12\%$ of the cosmic sulfur abundance. \par
Two other sulfur-bearing species have been detected on the grains to date: SO$_2$ and OCS \citep{Geballe1985,Palumbo1995,Boogert1997}. These are respectively rated as possible and likely detections in the review of \citet{Boogert2015}. However, the measured abundance of solid OCS by \citet{Palumbo1997} would only account for $\sim0.5\%$ of the cosmic sulfur abundance and SO$_2$ has an upper limit of 6\%. These observations are of high mass young stellar objects and the chemistry may be different to that found in outflows. In fact, in modelling work done to explain the sulfur ion abundances in L1157-B1, OCS was required to be a major carrier of sulfur on the grains \citep{Podio2014}.\par
In this work, observations of five sulfur-bearing species are presented. Multiple transitions of CCS, H$_2$CS, OCS, SO and SO$_2$ have been observed along with isotopologues of SO and SO$_2$. In Section~\ref{sec:observe}, details of the observation and data reduction are given. The column density of these species are measured and the properties of the emitting gas are discussed in Section~\ref{sec:results}. Finally, the conclusion is given in Section~\ref{sec:conc}.
\section{Observations \& Processing}
\label{sec:observe}
\subsection{IRAM-30m Observations}
The data presented here were collected as part of the IRAM-30m large programme ASAI \citep{Lefloch2018}, which included a systematic search for lines of sulfur-bearing species between 80 and 350 GHz. The data includes over 140 detected transitions of five species: CCS, H$_2$CS, OCS, SO, SO$_2$ along with the $^{34}$S isotopologues of SO and SO$_2$. These were obtained observing the L1157-B1 bowshock, with pointed co-ordinates $\alpha _{J2000} =20^h39^m10.2^s$, $\delta _{J2000} =+$\ang{68;01;10.5}, which is offset by $\Delta\alpha=+25".6 $, $\Delta\delta=-63".5$ from the position of L1157-mm. The beam size varied from 30$"$ at $\sim$80 GHz to 7$"$ at $\sim$ 340 GHz. Pointing was monitored using NGC7538 and found to be stable, corrections were typically less than 3".\par
The data were obtained using the IRAM-30m telescope's EMIR receivers with the Fourier Transform Spectrometer (FTS). This gave a spectral resolution of 200 kHz which corresponds to velocity resolutions between 0.7 and \SI{2.2}{\kilo\metre\per\second}. Intensities are expressed in units of antenna temperature corrected for atmospheric absorption and sky coupling (T$_{A}^*$). Where intensities are given in units of main-beam brightness temperature (T$_{MB}$), the efficiencies required for this conversion (B$_{eff}$ and F$_{eff}$) were interpolated from the values given in Table 2 of \citet{kramer2013} which can be found at \url{http://www.iram.es/IRAMES/mainWiki/Iram30mEfficiencies}. A nominal 20\% calibration uncertainty is assumed and propagated to values such as the integrated emission.
\subsection{Line Identification \& Properties}
\label{sec:lineprops}
Lines were identified by comparing the frequencies of emission peaks to the JPL catalog \citep{pickett1998} accessed via Splatalogue\footnote{\url{http://www.splatalogue.net/}}. Baselines were removed using the GILDAS CLASS software package\footnote{\url{http://www.iram.fr/IRAMFR/GILDAS}} and the rms noise level was calculated for every spectrum by considering velocity ranges of $\pm$\SI{50}{\kilo\metre\per\second} around the detected transitions. The list of identified lines is given in Tables~\ref{table:lines} to \ref{table:lines3} in Appendix~\ref{appendix:lines}. The spectra are shown in Figures~\ref{fig:ccsspectra} to \ref{fig:34so2spectra} in Appendix~\ref{appendix:profiles}. Any line that was blended or contaminated with another molecular line has been removed from the data set. \par
From the line profiles of the detected transitions, two different velocity regimes were identified: a moderate velocity regime (between V$_{min}$ = -8 \si{\kilo\metre\per\second} and V$_{max}$ = 6 \si{\kilo\metre\per\second}) and a high velocity regime (between V$_{min}$ = -20 \si{\kilo\metre\per\second} and V$_{max}$ = -8 \si{\kilo\metre\per\second}). All molecules show emission for the moderate velocity regime and the integrated emission reported in Table~\ref{table:lines} was measured between those limits. The lines peak at approximately \SI{0}{\kilo\metre\per\second}, to within the velocity resolution of the spectra. This is blue shifted with respect to the systematic velocity of \SI{2.6}{\kilo\metre\per\second} \citep{Bachiller1997} which was also found to be the case for other species in L1157-B1 including CS \citep{Codella2010,Gomez-Ruiz2015}. For the high velocity regime, many SO and SO$_2$ transitions show emission above the 3$\sigma$ noise level. In fact, both species have many transitions where the line profiles present secondary peaks within the high velocity range. The 278.887 GHz transition of H$_2$CS also shows significant emission in this range. The high velocity emission is analysed separately and is discussed in Section~\ref{sec:sosecondary}. In Table~\ref{table:lines}, the minimum and maximum terminal velocity of the individual line profiles, measured considering emission above a 3$\sigma$ noise level, are also reported. Finally, no emission with velocities more blue-shifted than \SI{-20}{\kilo\metre\per\second} are detected in the spectra of any sulfur-bearing species considered in this work. This indicates that these species do not likely participate of the high-excitation CO velocity component reported by \citet{lefloch2012}.\par
\subsection{Rotation Diagram Analysis}
The rotational temperatures and column densities of each species were calculated through use of rotation diagrams. For the rotational diagrams, the upper state number density for each transition was calculated from the integrated emission using,
\begin{equation}
\label{Nu}
N_u = \frac{8\pi k\nu^2}{hc^3A_{ul} \eta_{ff}} \int T_{MB}dv,
\end{equation}
where $\eta_{ff}$ is the filling factor. The spectroscopic parameters are taken from the JPL catalog \citep{pickett1998}. From plots of ln(N$_u$/g$_u$) against $E_u$, the column density and rotational temperature of each species can be found.\par
This technique relies on the assumption that the emission is optically thin and the gas and radiation are in LTE. In this work, it is first assumed that the emission is optically thin for all transitions of each species. However, as explained in Section~\ref{sec:nonlte}, for those molecules for which collisional coefficients are available (H$_2$CS, OCS, SO and SO$_2$), we also perform a non-LTE analysis. As shown in Section~\ref{sec:radexresults}, a good agreement is found between the two methods.\par
The same source size is used for all species and transitions and is taken to be 20". This is estimated from interferometric CS maps of the region taken by \citet{Benedettini2007}. The filling factor is calculated as,
\begin{equation}
\label{ff}
\eta_{ff}= \frac{\theta_S^2}{\theta_{MB}^2+\theta_S^2}
\end{equation}
where $\theta_{MB}$ and $\theta_{S}$ are the beam and source sizes respectively. The beam size is derived from the frequency using the formula $\theta_{MB}=$2460"/frequency(GHz) \citep{kramer2013}.
\subsection{Non-LTE Analysis}
\label{sec:nonlte}
Where collisional coefficients were available, the radiative transfer code RADEX \citep{vandertak2007} was used to estimate the column density of each molecule and the properties of the emitting gas. RADEX assumes a uniform medium and treats optical depth effects through an escape probability that is dependent on the assumed geometry. A slab geometry is used in this work. The species H$_2$CS, OCS, SO and SO$_2$ have collisional data available in the LAMDA database\footnote{\moldataurl}\citep{Schoier2005} and so these were fit with RADEX. This represents an improvement over the rotation diagram analysis as LTE and optically thin emission no longer need to be assumed.\par
RADEX assumes that the source fills the beam which is unlikely to be the case for the species reported here. Therefore, the flux of each transition was adjusted by a filling factor in the same way as the rotation diagram analysis (i.e. by assuming a source size of 20"). Note, however, the column density values derived from the RADEX fits change by up to a factor of 2 if a smaller souce size of 10" or an extended source is assumed and are often within the reported error bars.\par
RADEX fits are described by three parameters: the gas density, gas temperature and species column density. A Bayesian approach to inferring these parameter values is taken. The posterior probability distributions of the parameters is given by Bayes' theorem:
\begin{equation}
P(\boldsymbol{\theta}|\boldsymbol{d}) = \frac{\mathcal{L}(\boldsymbol{d}|\boldsymbol{\theta})P(\boldsymbol{\theta})}{P(\boldsymbol{d})} \propto \mathcal{L}(\boldsymbol{d}|\boldsymbol{\theta})P(\boldsymbol{\theta})
\end{equation}
where $\boldsymbol{\theta}$ represents the parameters $n_H$, $T_{kin}$ and N. $P(\boldsymbol{\theta})$ is the prior probability distribution of the parameters, representing any previous knowledge of the parameter values. $P(\boldsymbol{d})$ is referred to as the Bayesian evidence but can be simply considered to be a normalizing factor for this work. $\mathcal{L}(\boldsymbol{d}|\boldsymbol{\theta})$ is the likelihood of the data given the parameters. This is related to the $\chi^2$ value through the relation $\mathcal{L}(\boldsymbol{d}|\boldsymbol{\theta})=exp(-\chi^2/2)$. In which,
\begin{equation}
\chi^2=\sum_i \left(\frac{F_{i,RADEX}-F_{i,obs} } { \sigma_{i,obs} }\right)^2
\end{equation}
where F$_{i,RADEX}$ and F$_{i,obs}$ are the RADEX predicted and observed fluxes of transition i respectively and $\sigma_{i,obs}$ is the uncertainty on the observed flux.\par 
To sample the posterior distribution in this work, the python package PyMultiNest \citep{Buchner2014} was used. This is a package for nested sampling, an algorithm in which the parameter space is sampled according to the prior distribution rather than in a random walk \citep{Skilling2004}. A number of samples is drawn from the prior distribution and their likelihoods are evaluated. The least likely samples are then replaced with more likely samples until the total probability density left unexplored is negligible. This has advantages over approaches such as the Metropolis-Hastings algorithm when the posterior distribution of the parameters is likely to be multimodal.\par
The priors are assumed to be uniform and non-zero in the range 0-\SI{300}{\kelvin} for temperature and in the range \num{e6}-\SI{e16}{\per\centi\metre\squared} for the column density. The prior was uniform in log-space for values of the gas density between \num{e4} and \SI{e8}{\per\centi\metre\cubed} to prevent high densities from being unreasonably oversampled. This was not a concern for the species column density due to the fact the fits were so strongly dependent on the column density value. The line width was originally included as a free parameter but found not to affect the probability distributions of the other parameters. It was then fixed at \SI{6}{\kilo\metre\per\second} which is the average FWHM line width measured for all transitions and all species for the fits that are reported here.\par
For H$_2$CS, the transitions were separated into ortho and para H$_2$CS and fit separately. Both the o-H$_2$CS and the p-H$_2$CS species have seven detected transitions, which is sufficient to constrain their column density. For the other species, the integrated emission from all detected transitions was used to constrain the RADEX fits. That is 14 lines for OCS, 23 lines of SO and 32 lines of SO$_2$.\par
The final outputs of the nested sampling routine are the marginalized and joint probability distributions for each fitted parameter, these are shown in Appendix~\ref{appendix:mcmc}. The benefit of this sampling procedure over a simple grid of $\chi^2$ values is the improved sampling of areas of interest and a probability distribution that fully describes the likelihood of different parameter values in the model. The reported values of the gas density, temperature and species column density in Section~\ref{sec:results} are the median values of the marginalized probability distributions, this corresponds to the most likely value for well constrained parameters. The reported uncertainties represent the interval containing 67\% of the probability density in the posterior distributions.\par
\section{Results}
\label{sec:results}
In the sections below, the column density of each detected species derived from either the rotation diagram analysis or RADEX fitting is presented and discussed. The properties of the emitting gas given by the RADEX fits are also discussed. All of these results are summarized in Table~\ref{table:colDens} along with the corresponding values for sulfur-bearing species in L1157-B1 taken from the literature.\par
Fractional abundances are also reported in the table. These are derived by assuming a CO abundance of 10$^{-4}$ and comparing the species column density to the column density of CO in the region. The CO emission from L1157-B1 can be divided into multiple emitting components so it is not obvious which value of the CO column density should be adopted. A value of N(CO)=\SI{1.5\pm0.5e17}{\per\centi\metre\squared} is chosen since this range contains the column densities of CO measured for the low excitation emission of CO in the region \citep[see][]{lefloch2012}. This is justified by the fact that the gas properties derived from RADEX in Section~\ref{sec:radex} are incompatible with the high density (\SI{e6}{\per\centi\metre\cubed}) and temperature (\SI{200}{\kelvin}) found for the high excitation CO emission \citep{lefloch2012} which is undetected in our sulfur-bearing spectra (see Section~\ref{sec:lineprops}).
\begin{table*}
\centering
\caption{Column density and fractional abundances for all species, tabulated with gas temperature and density fits from RADEX}
\begin{tabular}{ccccc}
\hline
\textbf{Species} & \textbf{N} & \textbf{Fractional Abundance} & \boldmath$\mathrm{T_{kin}}$ & \boldmath$\mathrm{log(n_{H2})}$ \\ 
 & \textbf{cm$^{-2}$} & & \textbf{K} & \textbf{\si{\per\cm\cubed}} \\ 
\hline
CCS$^{(1)}$ & \num{1.2\pm0.7e13}& \num{7.8\pm5.6e-9}& \num{6.3\pm1.7}& - \\ 
CCS$^{(2)}$ & \num{2.8\pm1.9e12}& \num{1.9\pm1.4e-9}& \num{47.9\pm28.0}& - \\ 
o-H$_2$CS & \num{2.6\pm0.5e13}& \num{1.7\pm0.7e-8}& \num{177.2\pm124.3}& \num{4.9\pm0.7e0} \\ 
p-H$_2$CS & \num{7.4\pm0.8e12}& \num{4.9\pm1.7e-9}& \num{93.2\pm29.6}& \num{5.0\pm0.1e0} \\ 
OCS & \num{6.6\pm0.5e13}& \num{4.4\pm1.5e-8}& \num{46.8\pm3.4}& \textgreater \num{e5} \\ 
SO & \num{1.8\pm0.1e14}& \num{1.2\pm0.4e-7}& \num{17.9\pm0.9}& \textgreater \num{e6} \\ 
SO Secondary & \num{3.5\pm0.4e12}& \num{2.4\pm0.8e-9}& \textgreater 100 & \num{5.7\pm0.1e0} \\ 
$^{34}$SO$^{(3)}$ & \num{6.3\pm6.5e12}& \num{4.2\pm4.5e-9}& \num{20.0\pm11.3}& - \\ 
SO$_2$ & \num{9.3\pm0.8e13}& \num{6.2\pm2.1e-8}& \num{48.0\pm6.3}& \num{5.7\pm0.1e0} \\ 
\\ 
\hline
\\ 
 H$_2$S$^{(4)}$  & \num{6e13} & \num{6e-8} & - & - \\ 
 CS$^{(5)}$	& \num{8e13}	&	\num{8e-8} & 50 - 100 & \num{e5}-\num{e6} \\ 
 SO$^{+ (6)}$ & \num{7e11} &\num{8e-10} & - & -\\ 
 HCS$^{+ (6)}$ & \num{6e11}	& \num{7e-10} & 80 & \num{8e5}\\ 
 SiS$^{(7)}$ & \num{2e13} &	\num{2e-8} & - & -\\ 
 \hline
\end{tabular}
 \\ 1) Lower excitation component, the temperature given is rotational. 2) Higher excitation component, the temperature given is rotational 3) Higher excitation component, the temperature given is rotational 4) \citet{Holdship2017erratum}. 5) \citet{Gomez-Ruiz2016}. 6) \citet{Podio2014}. 7) \citet{Podio2017}.
 \\ 
\label{table:colDens}
\end{table*}
\subsection{Rotation Diagram Analysis}
\label{sec:rotdiagresults}
The results of the rotational diagram approach are considered first. It is often necessary to employ multiple LTE components when fitting emission in shocked regions \citep[e.g. CH$_3$OH,][]{Codella2010}. Indeed, many of the species presented here are better fit by two LTE components than by one, including CCS, OCS, SO and $^{34}$SO.\par
\begin{figure}
\includegraphics[width=0.5\textwidth]{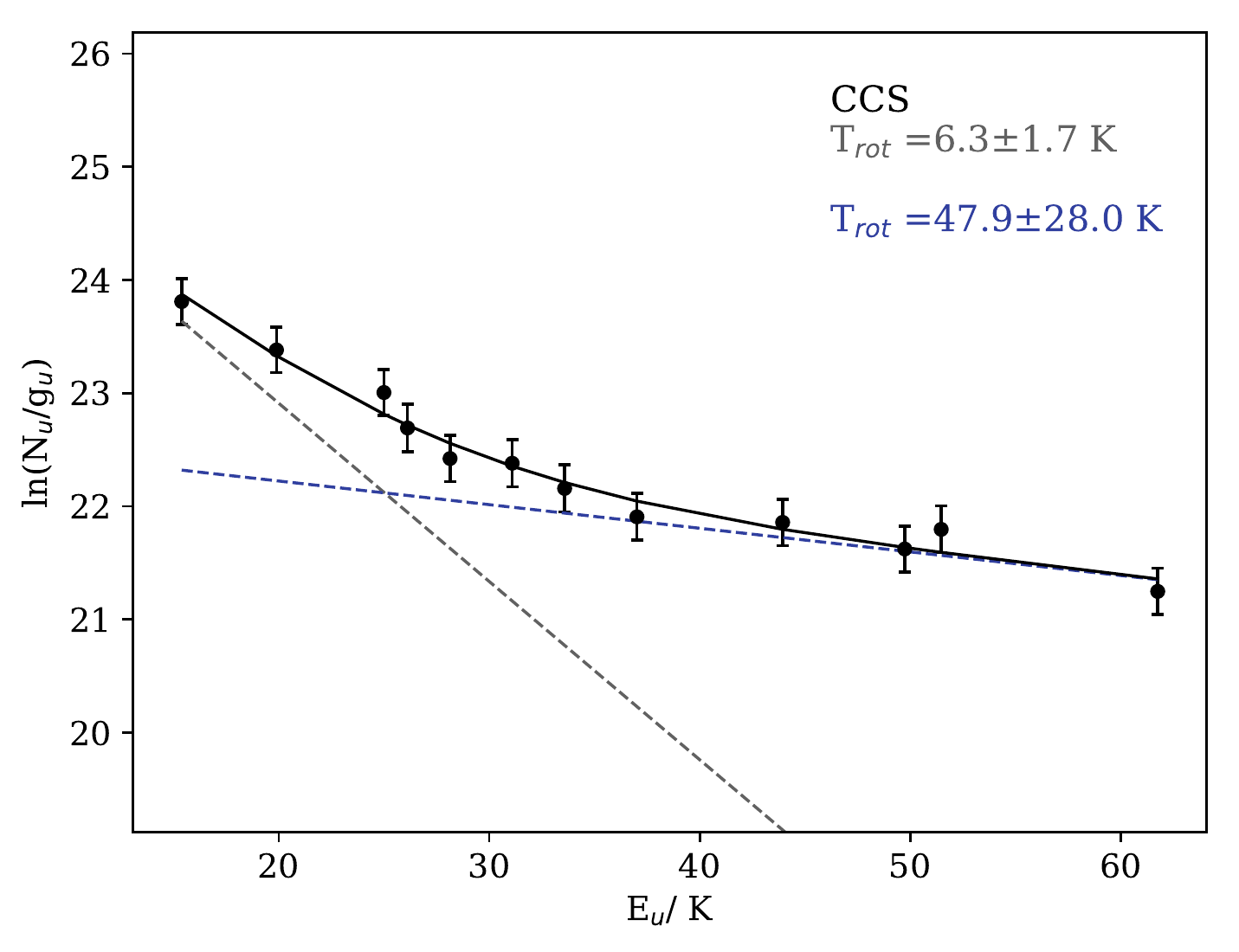}
\caption{Rotation Diagram for detected CCS transitions\label{fig:ccsrot}. Two components have been fit due to clear break in gradient at $E_u$=\SI{40}{\kelvin}. The black solid line shows the combined value of the two components.}
\end{figure}
In Figure~\ref{fig:ccsrot}, the rotation diagram of CCS is plotted. It shows that the CCS emission is well fit by a combination of two components, one with a rotation temperature of \SI{6}{\kelvin} and another \SI{48}{\kelvin} component. The column densities derived from these two components differ by a factor of 4.\par
\begin{figure}
\includegraphics[width=0.5\textwidth]{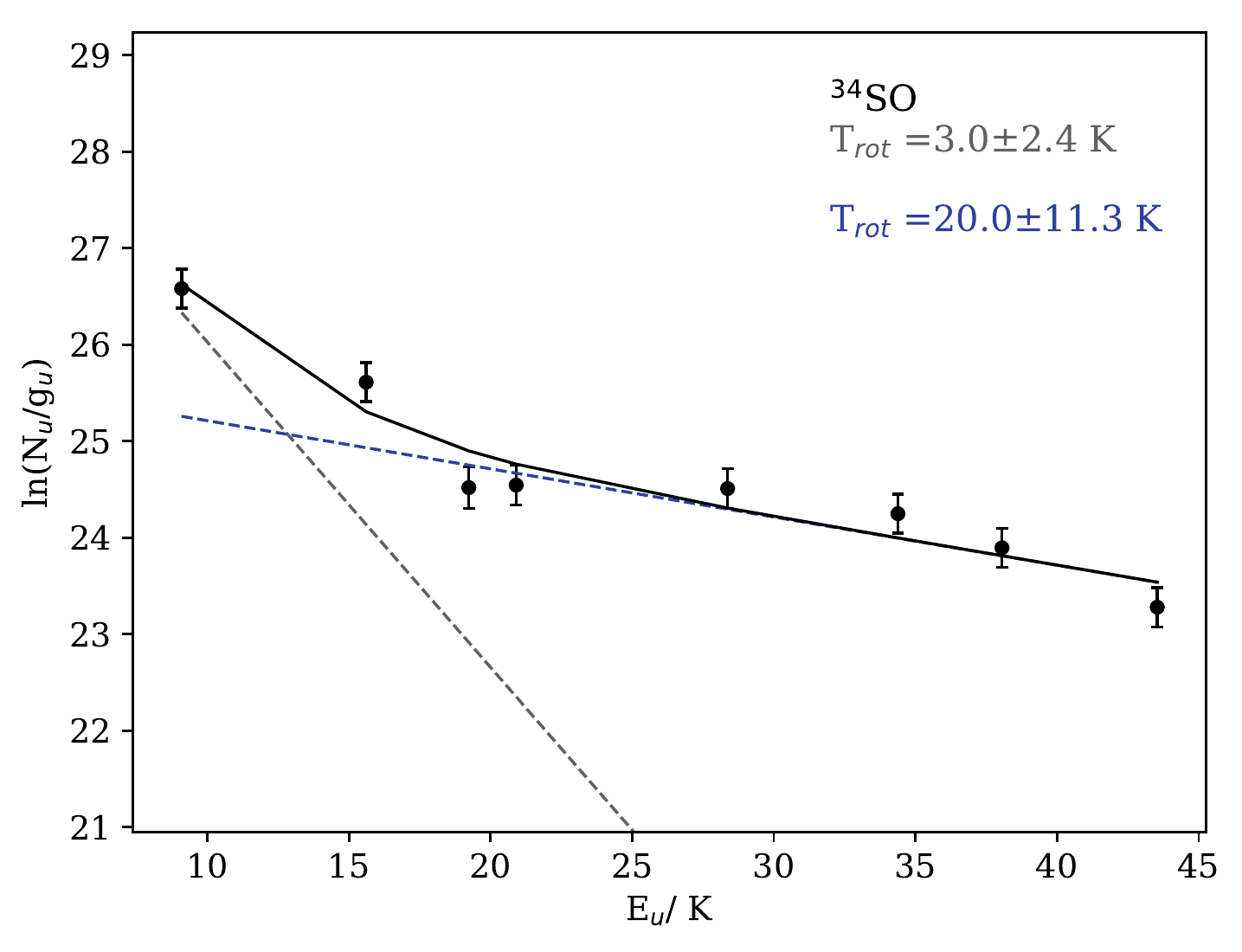}
\caption{Similar to Figure~\ref{fig:ccsrot} for detected $^{34}$SO transitions\label{fig:34sorot}. }
\end{figure}
\begin{figure}
\includegraphics[width=0.5\textwidth]{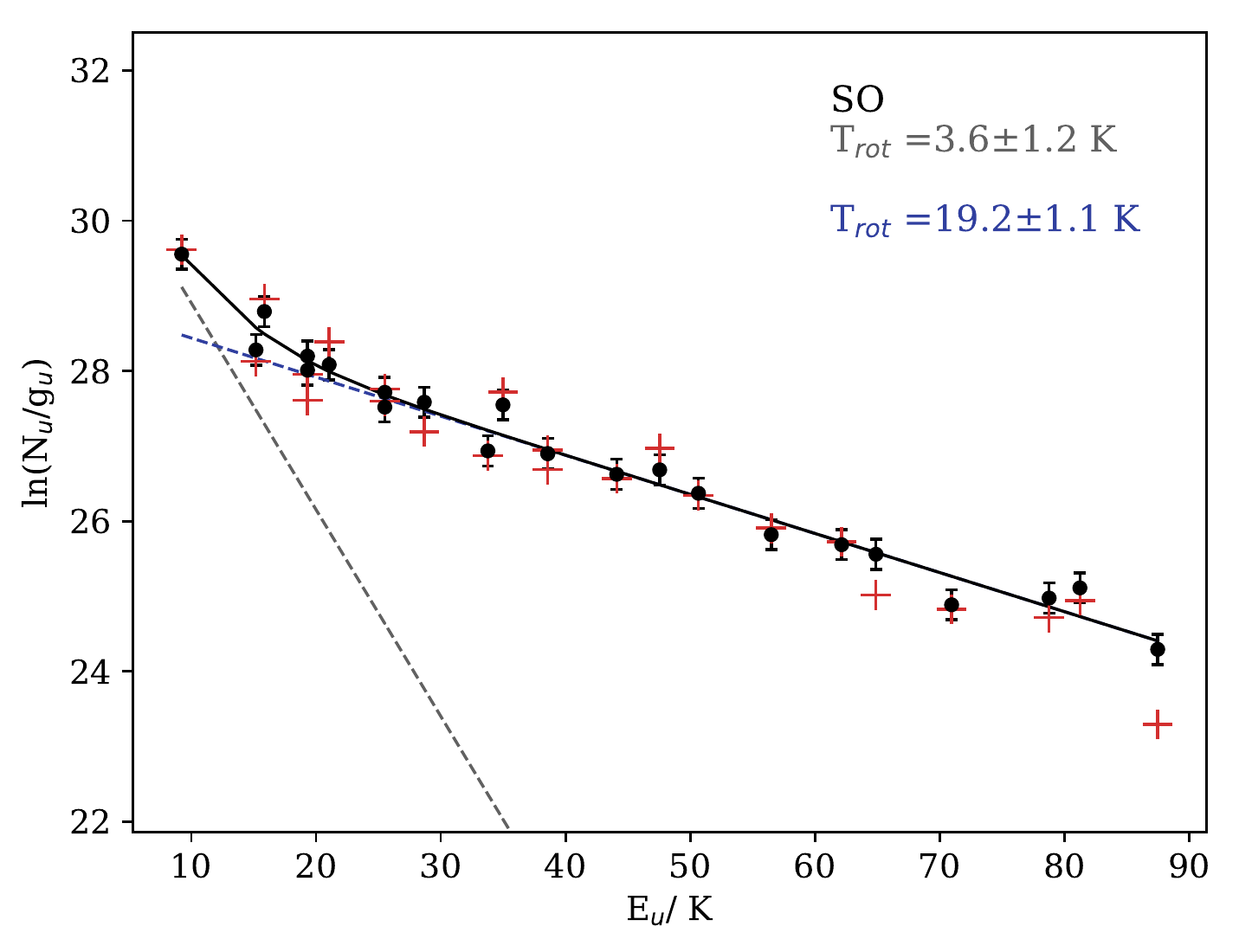}
\caption{Similar to Figure~\ref{fig:ccsrot} for SO. The best fit LTE components are remarkably similar to those found for $^{34}$SO. The overplotted red points indicate the upper state column density derived from the RADEX best fit fluxes to give an indication of the quality of the RADEX fit.\label{fig:sorot}}
\end{figure}

The rotation diagram of $^{34}$SO is presented in Figure~\ref{fig:34sorot} and shows a broken trend similar to CCS. This is also seen in Figure~\ref{fig:sorot}, where two components with rotational temperatures of $\sim$\SI{3}{\kelvin} and $\sim$\SI{20}{\kelvin} are required to fit the transitions of the main SO isotopologue. Although these are the lowest values of T$_{rot}$ derived from this sample of sulfur-bearing species, it should be noted that the same temperatures are independently inferred from $^{34}$SO. In addition, they are consistent with those measured from high-angular resolution NOEMA maps of SO toward the L1157-B1 cavity with T$_{rot}<$ \SI{10}{\kelvin} and T$_{rot}\sim$ \SI{24}{\kelvin} for, respectively, the low- and high-temperature components found at moderate velocities (Feng et al. in prep.). \par
Comparing the column densities of SO and $^{34}$SO, a ratio of \num{23.9} is obtained for the higher temperature component with T$_{rot}\sim$\SI{20}{\kelvin}. This is consistent with the $^{34}$S/$^{32}$S ratio measured terrestrially \citep[22.13][]{rosman1998} and implies the SO rotation diagrams do not strongly suffer from optical depth effects. This is further supported by the fact that comparing the integrated intensity of the strongest $^{34}$SO transitions with the equivalent transitions of the main isotopologue gives optical depths $\tau < 0.2$ assuming an isotopic ratio $^{32}$S/$^{34}$S of 22.13 \citep{rosman1998}.\par
\subsection{Non-LTE Analysis}
\label{sec:radexresults}
The most robust results from the RADEX fitting are the column densities of the species. The fitting procedure strongly constrained the column density in every case. The probability distributions from the sampling procedure are shown in Appendix~\ref{appendix:mcmc}. The values given in Table~\ref{table:colDens} are the median values from those probability distributions and the reported errors give the 67\% probability interval.\par
The most likely column density values of the ortho and para H$_2$CS species are the same regardless of whether they are fit separately with RADEX or are fit using the same gas density and temperature. If it is assumed that the two spin isomers of H$_2$CS trace the same gas, an ortho to para ratio can be calculated from their respective column densities. The best fit column density value for each species reported in Table~\ref{table:colDens} give an ortho to para ratio of 2.8$\pm$0.1 for H$_2$CS in L1157-B1. This is consistent with the statistical limit of 3. \par
\begin{figure}
\includegraphics[width=0.5\textwidth]{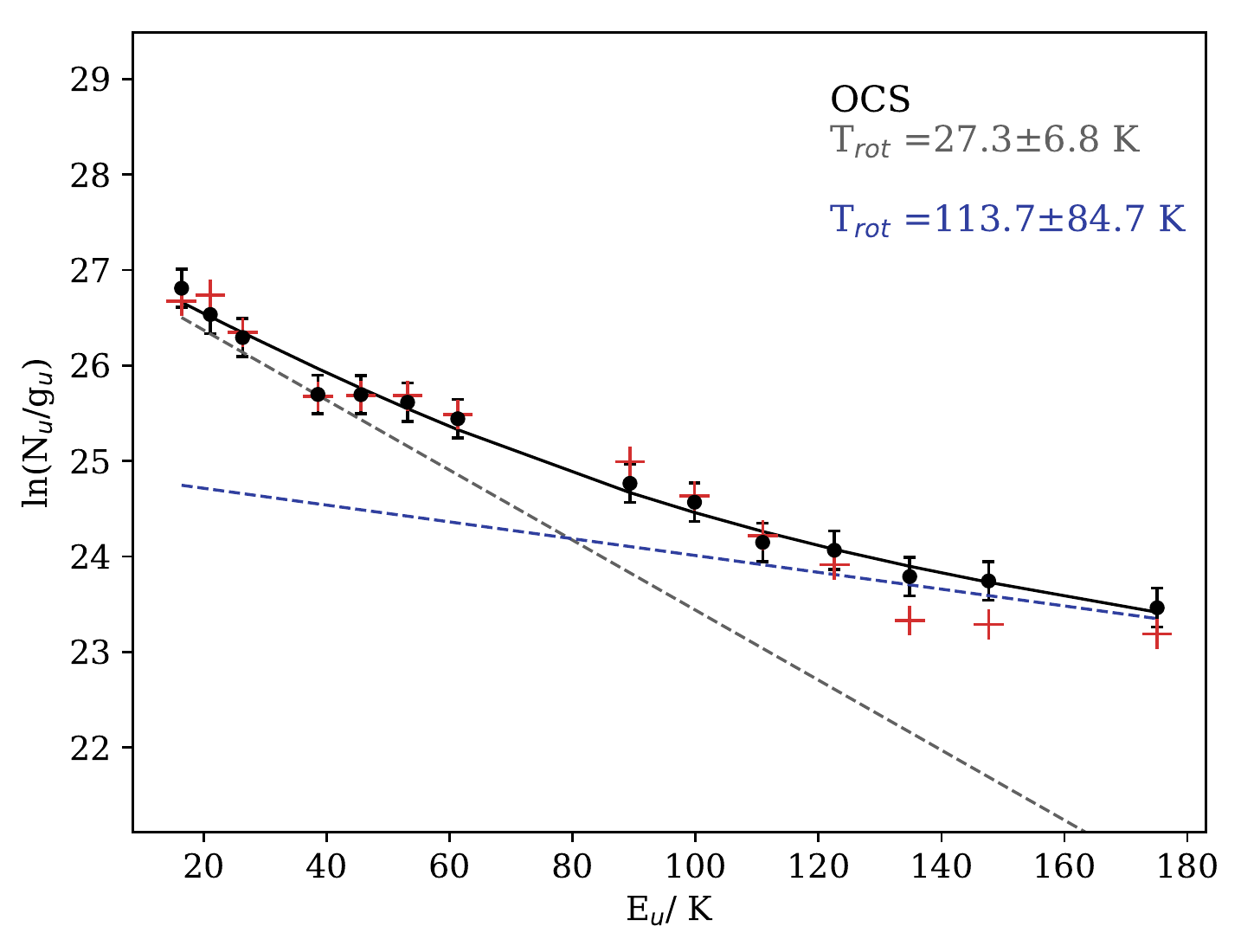}
\caption{Rotation Diagram for observed OCS transitions plotted in black. Again, two LTE components have been fitted and the black line shows the combined value. The overplotted red points indicate the column density derived from the RADEX best fit fluxes.\label{fig:ocsrot}}
\end{figure}
It was possible to strongly constrain the column density of OCS, SO and SO$_2$. Figures~\ref{fig:sorot} and \ref{fig:ocsrot} show the RADEX fluxes (see the red markers in the plots) for the most likely values from the OCS and SO fitting, converted to column densities through Equation~\ref{Nu} and plotted as rotation diagrams with the original data. This allows the quality of the rotation diagram analysis and RADEX fit for each species to be assessed visually. The RADEX fits are in good agreement with the data, although they start to systematically fall below the measured points at low $E_u$ because the RADEX calculations do not consider a two excitation component model.\par
\begin{figure}
\includegraphics[width=0.5\textwidth]{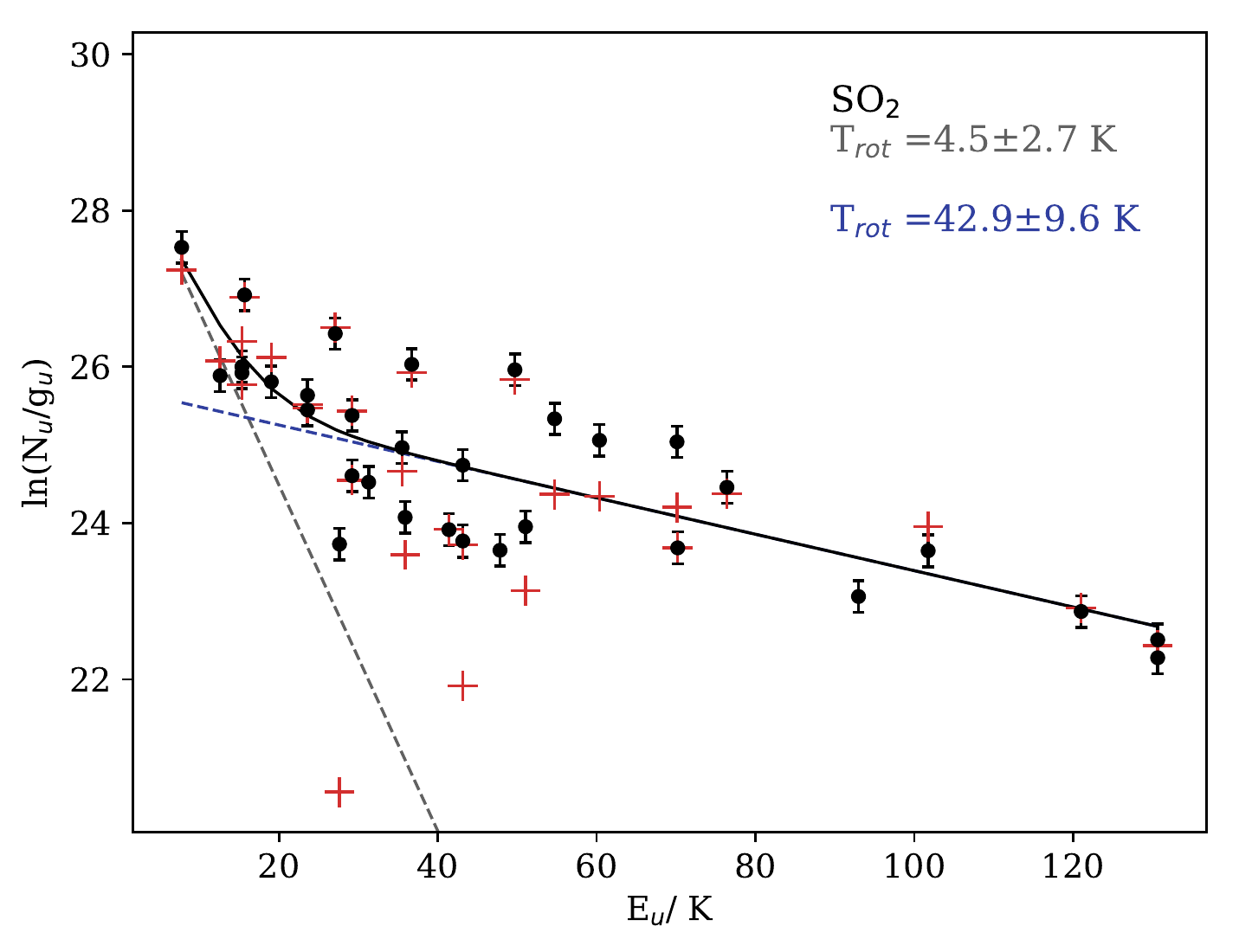}
\caption{Similar to Figure~\ref{fig:ocsrot} for observed SO$_2$ transitions. The large degree of scatter is typical of non-linear molecules due to the fact optical depth varies strongly between transitions for such molecules. \citep{goldsmith1999}.\label{fig:so2rot}}
\end{figure}

Figure~\ref{fig:so2rot} shows the best SO$_2$ RADEX fit, it gives good agreement with the lower $E_u$ transitions but fails at high $E_u$. It is likely that the SO$_2$ emission arises from multiple gas components and the high $E_u$ emission is from another, hotter gas component. This is consistent with SO$_2$ interferometry of the region (Feng et al. in prep.) which shows a variation in the excitation conditions of SO$_2$ across L1157-B1.\par%
\subsection{Bulk Gas Properties}
\label{sec:radex}
The physical properties of the emitting gas that best reproduce the observed line fluxes are derived from the RADEX fits to each species. These are tabulated in Table~\ref{table:colDens} and a more detailed view can be seen in Appendix~\ref{appendix:mcmc} where the probability distributions are plotted for each parameter.\par
\begin{figure}
\includegraphics[width=0.5\textwidth]{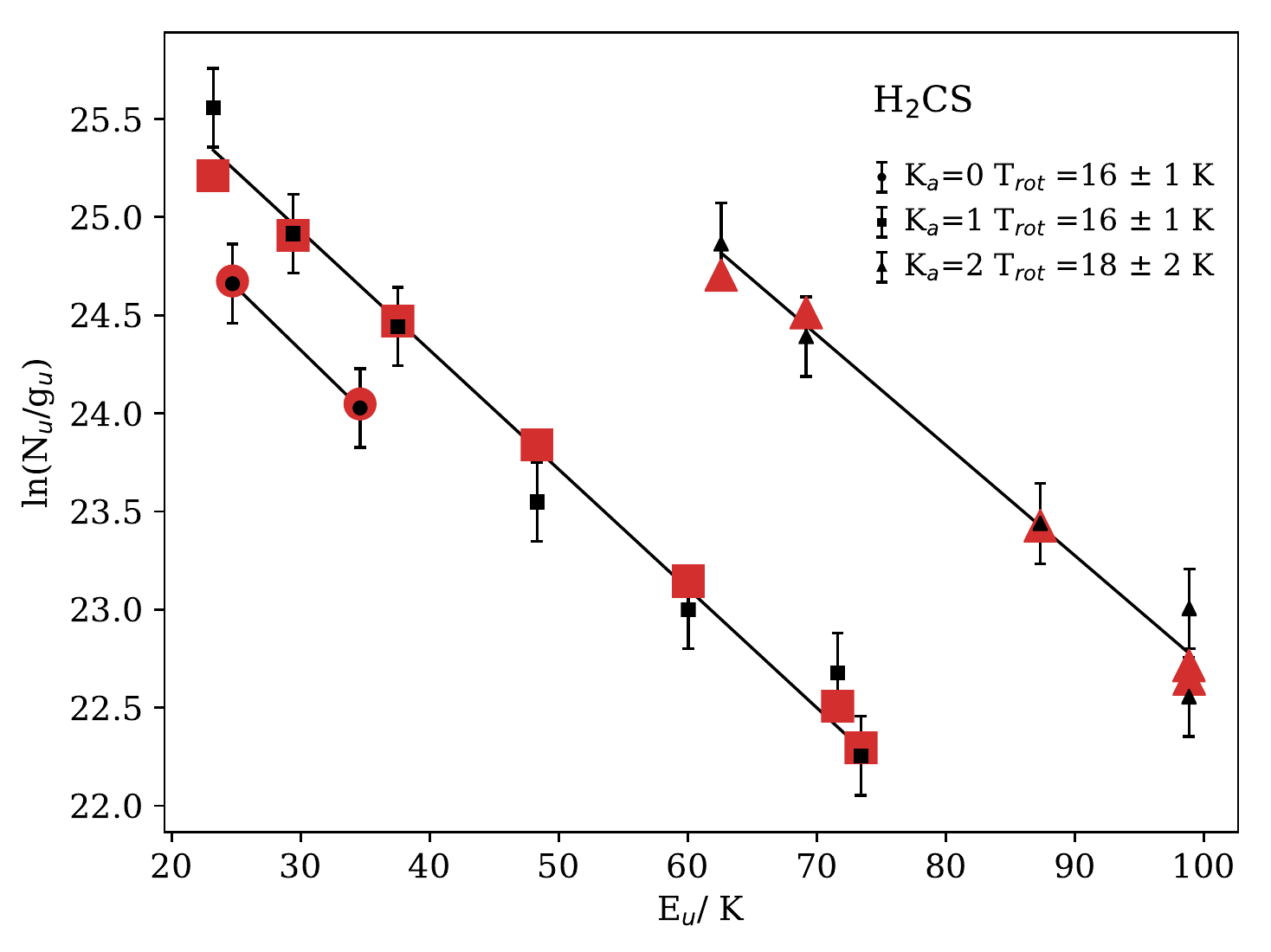}
\caption{Rotation Diagram for detected H$_2$CS transitions plotted in black with ortho transitions marked as squares and para transitions as circles and triangles. Three distinct ladders can be seen, each made up from transitions of the same K$_a$ quantum number. The equivalent red points indicate the values given by the best fit RADEX model. \label{fig:h2csrot}}
\end{figure}

The fits to the H$_2$CS spin isomers present an interesting case. A priori, one might expect the two spin isomers to trace the same gas. Indeed, both have strong peaks in the gas density probability distribution at approximately \SI{e5}{\per\centi\metre\cubed}. However, the o-H$_2$CS probability distributions of the temperature and density are highly degenerate and so higher density, low temperature solutions exist for that spin isomer. The p-H$_2$CS fits are better constrained because at low densities, H$_2$CS transitions form separate ladders in a rotation diagram in which transitions of the same K$_a$ quantum number follow separate linear trends \citep{Cuadrado2017}. This can be seen in Figure~\ref{fig:h2csrot} where the p-H$_2$CS transitions form two ladders. It is likely that the o-H$_2$CS data set simply does not cover a large enough K$_a$ range to give accurate fits to the gas properties.\par
If it is assumed that the H$_2$CS isotopologues do in fact trace the same gas, a RADEX fit can be made with one gas temperature and density for both species. In this case, the same column densities are obtained for each species and the gas properties obtained are the same as those obtained for the fits to p-H$_2$CS only. The gas temperature and density obtained for p-H$_2$CS is also consistent with those found for the L1157-B1 cavity from LVG fits to CS emission \citep{Gomez-Ruiz2015}.\par
The RADEX fits to the emission of OCS only gives lower limits on the gas density. This is likely to be because the level populations of the detected transitions of these species are thermalized and transitions with a broader range of A$_{ij}$ values are required to break the degeneracy. The gas temperature is well constrained although, at T$_{kin}=$\SI{46.8\pm3.4}{\kelvin}, it is somewhat lower than that found for p-H$_2$CS.\par 
SO fits also provide only a lower limit on the gas density. This lower limit is \SI{e6}{\per\centi\metre\cubed} and it is interesting to note the critical densities of the detected SO transitions are no larger than \SI{e6}{\per\centi\metre\cubed}. Fits to the SO emission also give the lowest gas temperature found for any species at T$_{kin}=$\SI{17.9\pm0.9}{\kelvin}. This is consistent with the rotational temperature inferred from the rotation diagram analysis and with values measured from interferometric data (see Section\ref{sec:rotdiagresults} and Feng et al. in prep).\par
Alternatively, it is possible that the low T$_{rot}$ SO emission does not arise from the shocked B1 cavity itself but from cooler, more extended gas. CO \citep{lefloch2012} and CS \cite{Gomez-Ruiz2015} observations taken as part of the ASAI survey show a component of extended, low excitation emission in addition to that from the B1 cavity with a temperature similar to that found for SO ($\sim$\SI{23}{\kelvin}). To check for consistency, a RADEX fit was performed using a filling factor of 1 due to the extended nature of the cold component and similar gas properties were recovered. \par
Finally, the RADEX fits to SO$_2$ are the most well constrained. A most likely temperature of T$_{kin}=$\SI{48.0\pm6.3}{\kelvin} is obtained which is consistent with the temperature found for OCS. The gas density value of n$_{H2}$=\SI{7.9e5}{\per\cm\cubed} is higher than those found for OCS or p-H$_2$CS but is still within the range found for the L1157-B1 cavity from LVG analysis of CS emission \citep{Gomez-Ruiz2015}.\par
Overall, the sulfur bearing species observed towards L1157-B1 appear to trace gas with properties that are consistent with those found previously in the region. Although there is some variation between species, they broadly trace warm (40-100 \si{\kelvin}) gas with a density between \SI{e5}{\per\cm\cubed} and \SI{e6}{\per\cm\cubed}. The exception is SO which traces much cooler gas. Such variation is to be expected considering that a complex shocked region is being fit with a single gas component and chemical effects may alter the distribution of each species.
\subsection{High Velocity Emission}
\label{sec:sosecondary}
A secondary peak, emitting in the velocity range V$_{min}$ = -20 \si{\kilo\metre\per\second} and V$_{max}$ = -8 \si{\kilo\metre\per\second} is evident in many SO lines and in a smaller number of SO$_2$ line. The presence of this peak in multiple transitions implies it is not due to contamination from other lines or species and that it comes from a weakly emitting, more blue shifted part of the bow shock. Furthermore, secondary peaks were observed in HCO$^+$ by \citet{Podio2014} so this is not unique to SO and SO$_2$. In fact, \citet{Benedettini2013} showed B1 to be made up of substructures and one such ``clump'' B1a showed secondary high velocity emission at \SI{-12}{\kilo\metre\per\second}. This coincides with the peak of the secondary emission.\par
\begin{figure}
\includegraphics[width=0.5\textwidth]{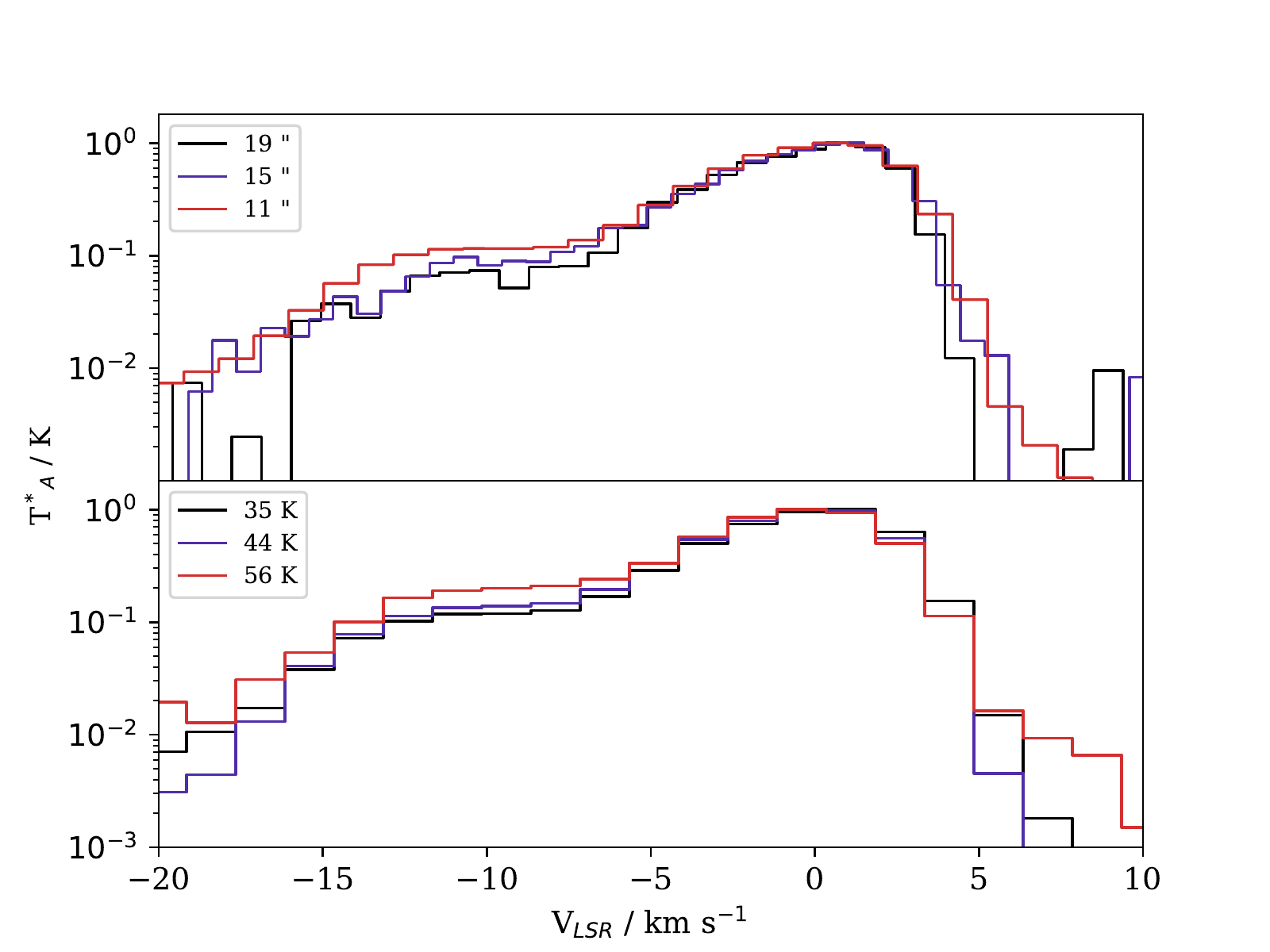}
\caption{Normalized and resampled line profiles of SO where the secondary emission peak at \SI{-12}{\kilo\metre\per\second} can be seen. The upper panel shows transitions with similar excitation properties but different frequencies, to illustrate the effect of beam size. The lower panel shows the smaller effect of excitation, with three transitions of similar frequency but differing $E_u$ plotted.\label{fig:sobump}}
\end{figure}

An example of the SO line profiles can be seen in the upper panel of Figure~\ref{fig:sobump}. Overplotted are the transitions of SO at 129.138, 158.971 and 219.949 GHz. Each has an $E_u$ in the range 26 to 25 K. With the peaks normalized, it is clear the bump emission is a larger fraction of the peak emission for smaller beams. This trend is clearly visible across all transitions of SO. This would imply that the secondary bump is closely centred on the pointed position of the telescope, as the relative emission of the bump decreases as the emission is averaged over a larger area.\par
Due to the frequency spacing of the SO transitions, it is hard to deconvolve the effect of the beam size from any excitation effects. However, small groups of transitions with varying excitation properties and similar (within 1") beam sizes can be compared and there is some evidence that the bump to peak ratio increases for higher excitation transitions. The lower panel of Figure~\ref{fig:sobump} demonstrates this for three transitions, each with a beam size of 11" or 10". Whilst not definitively shown by the spectra, this conclusion is supported by the RADEX analysis below.\par
The pointed co-ordinates for the SO observations are extremely close to the B1a peak seen in CS and the bump is strongest with small beam sizes, consistent with the observational result that the B1a clump is no more than 8" \citep{Benedettini2013}. These points and the fact the secondary emission peaks at the same velocity as the high velocity emission associated with B1a makes it likely that the secondary SO emission originates from the B1a substructure rather than B1 as a whole.
The SO and SO$_2$ spectra were then integrated between \SI{-8}{\kilo\metre\per\second} and \SI{-20}{\kilo\metre\per\second}. This gave an estimate for the total flux in the secondary peak for each transition, which was then analysed with RADEX in the same way as the main emission. Considering the likely identification of the secondary emission as belonging to the B1a clump, a source size of 7" was assumed, the size of B1a used for LVG fitting of the CS emission in \citet{Benedettini2013}.\par
RADEX fits to the high velocity SO emission favour lower densities than the lower limit found for the main peak emission. For the high velocity emission, the most likely value is n$_{H2}$=\SI{5.0e5}{\per\cm\cubed} with a 1$\sigma$ interval of \SI{3.9e5}{\per\cm\cubed} to \SI{7.4e5}{\per\cm\cubed} whereas for the lower limit for the main peak is \SI{e6}{\per\cm\cubed}. The secondary bump also has a higher temperature than the main peak. It is not well constrained but has a lower limit of \SI{100}{\kelvin}. This value is consistent with the temperature range inferred by \citet[][from 53 to 132 K]{Benedettini2013}. The secondary SO emission has a column density of \SI{5.0\pm0.3e12}{\per\centi\metre\squared}, a factor of $\sim$ 40 less than the bulk of the SO emission.\par
The gas properties derived from RADEX fits to the high velocity SO emission gas are consistent with the LVG modelling of high velocity CS emission from the B1a clump from \citet{Benedettini2013}. That work similarly found a temperature varying from a few tens to a few hundred Kelvin and a density of up to \SI{5e5}{\per\cm\cubed}. In fact, the $\chi^2$ results presented in Figure 7 of that paper are similar to the joint probability distribution for gas temperature and density derived from the secondary SO emission shown in Appendix~\ref{appendix:mcmc}.\par
Very few SO$_2$ transitions show secondary emission that is clearly above the spectrum rms. However, the flux of each spectrum in the high velocity range was extracted and fit with RADEX. It was not possible to constrain the density or temperature of the gas. However, as has been the case for all sulfur bearing species, a best fit column density independent of the gas conditions was found. If the high velocity emission of SO$_2$ comes from the B1a clump, it has an SO$_2$ column density of \SI{8.1\pm2.0e11}{\per\centi\metre\squared}.\par
Finally, the column density associated with the high velocity H$_2$CS emission was also estimated. Only one H$_2$CS transition shows significant emission (ie. above the 3 sigma level) in the high velocity regime. Therefore, to find the column density with RADEX, the density and temperatures obtained for the high velocity SO emission were used. Incorporating the uncertainty of the results from SO and the uncertainty on the integrated emission of the H$_2$CS 278.887 GHz transition a column density between \SI{e11}{\per\centi\metre\squared} and \SI{e12}{\per\centi\metre\squared} was found for H$_2$CS in the high velocity regime.\par
\section{Conclusion}
\label{sec:conc}
Observations of CCS, H$_2$CS, OCS, SO and SO$_2$ towards L1157-B1 have been presented. RADEX fits have been made to the detected emission from each species, constraining the column densities of the species and the temperature of the gas they trace. The fits do not always strongly constrain the gas density of the emitting region but the fits to p-H$_2$CS and SO$_2$ give gas densities in the range of those found previously for L1157-B1. \par
The column densities of all five species are reported, together with a summary of the column densities of all of the sulfur-bearing species detected in L1157-B1 by the ASAI collaboration. These have been converted to fractional abundances using the CO column density from \citet{lefloch2012}. The sum of the abundances of these species accounts for approximately 10\% of the total sulfur budget, assuming a solar sulfur abundance.\par
Interferometry of the region shows clumpy substructure to L1157-B1. This is evident in the SO spectra, which show secondary emission at higher velocities than the main peak. The peak velocity of this secondary emission (\SI{-12}{\kilo\metre\per\second}), the RADEX derived properties and increasing brightness with smaller beams indicate that this emission comes from the "high speed bullet" associated with the B1a clump \citep{Benedettini2013}. From this identification, the gas properties of B1a can be constrained using RADEX fits to the SO emission.\par
\section{Acknowledgements}
J.H. was funded by an STFC studentship (ST/M503873/1) and STFC grant (ST/M001334/1). L.P. has received funding from the European Union Seventh Framework Programme (FP7/2007-2013) under grant agreement No. 267251. I.J.-S. acknowledges the financial support received from the STFC through an Ernest Rutherford Fellowship (proposal number ST/L004801). MT and RB acknowledge support from MINECO project AYA2016-79006-P. This work has been supported by the project PRIN-INAF 2016 The Cradle of Life - GENESIS-SKA (General Conditions in Early Planetary Systems for the rise of life with SKA). We thank the anonymous referee for their comments which improved this manuscript.
\software{GILDAS/CLASS (\url{http://www.iram.fr/IRAMFR/GILDAS}), RADEX \citep{vandertak2007}, PyMultiNest \citep{Buchner2014}}



\bibliographystyle{mnras}
\bibliography{sulfur}



\appendix
\section{Line Properties}
\label{appendix:lines}
Tables~\ref{table:lines} to \ref{table:lines3} give the spectroscopic and measured properties of the detected line. Spectroscopic properties are taken from the JPL catalog \citep{pickett1998} via splatalogue (http://www.cv.nrao.edu/php/splat/).
\begin{table*}
\caption{Line properties of detected transitions including: upper state energy (E$_u$); Einstein coefficients (A$_{ij}$); beam size ($\Theta_{MB}$); the peak antenna temperature with the spectrum rms in brackets; the peak, minimum and maximum velocities of the spectra, the velocity resolution, the forward and beam efficiencies, the filling factor ($\eta_{ff}$), and integrated emission ($\int T_{MB} dv$) measured in the moderate velocity regime discussed in Section~\ref{sec:lineprops}.}
\begin{tabular}{cccccccccccc}
\hline
 \textbf{Freq} & \textbf{Transition} & \boldmath$\mathrm{E_u}$  & \boldmath$\mathrm{log(A_{ij})}$ & \boldmath$\mathrm{\Theta_{MB}}$ & \boldmath$\mathrm{T_{A*,peak}}$ & \boldmath$\mathrm{V_{peak}}$ & \boldmath$\mathrm{V_{min}/V_{max}}$ &  \boldmath$\mathrm{\Delta V}$ & \boldmath$\mathrm{F_{eff}/B_{eff}}$ & \boldmath$\mathrm{\eta_{ff}}$ & \boldmath$\mathrm{\int T_{MB} dv}$ \\ 
 \textbf{GHz} & & \textbf{K} & & \textbf{\si{\arcsecond}} & \textbf{mK} &\textbf{km s$^{-1}$} & \textbf{km s$^{-1}$} & \textbf{km s$^{-1}$} & & & \textbf{K km s$^{-1}$} \\ 
\hline 
 &&&&&\textbf{CCS  ($^3\Sigma^-$)}&&\\ 
 \hline 
81.505 &N=6-5,J=7-6 & 15.4 & -4.61 & 30 & 33 (2) & -0.3 & -6.0/2.6 & 1.4 & 1.0& 0.3& 0.19 (0.04) \\ 
90.686 &N=7-6,J=7-6 & 26.1 & -4.48 & 27 & 11 (2) & 0.0 & -2.6/2.6 & 1.3 & 1.2& 0.3& 0.08 (0.02) \\ 
93.870 &N=7-6,J=8-7 & 19.9 & -4.42 & 26 & 31 (1) & 0.1 & -4.9/2.6 & 1.2 & 1.2& 0.4& 0.20 (0.04) \\ 
99.866 &N=8-7,J=7-6 & 28.1 & -4.35 & 25 & 11 (1) & -0.9 & -3.3/1.4 & 1.2 & 1.2& 0.4& 0.07 (0.02) \\ 
103.640 &N=8-7,J=8-7 & 31.1 & -4.30 & 24 & 10 (3) & -0.8 & -0.8/2.6 & 1.1 & 1.2& 0.4& 0.09 (0.02) \\ 
106.347 &N=8-7,J=9-8 & 25.0 & -4.25 & 23 & 31 (2) & 0.4 & -5.1/2.6 & 1.1 & 1.2& 0.4& 0.20 (0.04) \\ 
113.410 &N=9-8,J=8-7 & 33.6 & -4.18 & 22 & 16 (2) & 0.5 & -2.6/1.6 & 1.0 & 1.2& 0.5& 0.09 (0.02) \\ 
131.551 &N=10-9,J=11-10 & 37.0 & -3.97 & 19 & 24 (4) & -0.1 & -1.0/1.7 & 0.9 & 1.2& 0.5& 0.13 (0.03) \\ 
142.501 &N=11-10,J=11-10 & 49.7 & -3.87 & 17 & 16 (4) & 0.1 & 0.1/1.0 & 0.8 & 1.2& 0.6& 0.11 (0.02) \\ 
144.244 &N=11-10,J=12-11 & 43.9 & -3.85 & 17 & 21 (4) & -1.4 & -2.3/1.8 & 0.8 & 1.2& 0.6& 0.16 (0.03) \\ 
156.981 &N=12-11,J=13-12 & 51.5 & -3.74 & 16 & 25 (5) & -1.1 & -2.6/1.9 & 0.7 & 1.3& 0.6& 0.19 (0.04) \\ 
166.662 &N=13-12,J=12-11 & 61.8 & -3.66 & 15 & 24 (5) & -0.2 & -0.9/0.5 & 0.7 & 1.3& 0.7& 0.11 (0.02) \\ 
\hline 
 &&&&&\textbf{o-H$_2$CS ($^1A_1$)}&&\\ 
 \hline 
104.617 &3(1,2)-2(1,1) & 23.2 & -4.86 & 24 & 85 (2) & -0.8 & -7.5/4.8 & 1.1 & 1.2& 0.4& 0.72 (0.14) \\ 
135.298 &4(1,4)-3(1,3) & 29.4 & -4.49 & 18 & 112 (3) & 0.0 & -6.9/3.5 & 0.9 & 1.2& 0.6& 0.89 (0.18) \\ 
169.114 &5(1,5)-4(1,4) & 37.5 & -4.18 & 15 & 122 (5) & -0.9 & -5.7/3.3 & 0.7 & 1.3& 0.7& 1.07 (0.21) \\ 
209.200 &6(1,5)-5(1,4) & 48.3 & -3.89 & 12 & 76 (3) & -0.8 & -6.4/2.6 & 1.1 & 1.5& 0.7& 0.74 (0.15) \\ 
244.048 &7(1,6)-6(1,5) & 60.0 & -3.68 & 10 & 58 (3) & -1.2 & -7.0/3.6 & 1.0 & 1.6& 0.8& 0.63 (0.13) \\ 
270.521 &8(1,8)-7(1,7) & 71.6 & -3.54 & 9 & 39 (5) & -1.4 & -3.6/3.0 & 2.2 & 1.7& 0.8& 0.61 (0.12) \\ 
278.887 &8(1,7)-7(1,6) & 73.4 & -3.50 & 9 & 30 (4) & -1.3 & -3.5/0.8 & 2.1 & 1.8& 0.8& 0.41 (0.08) \\ 
\hline 
 &&&&&\textbf{p-H$_2$CS ($^1A_1$)}&&\\ 
 \hline 
103.051 &3(2,1)-2(2,0) & 62.6 & -5.08 & 24 & 10 (2) & 0.3 & -3.1/0.3 & 1.1 & 1.2& 0.4& 0.07 (0.02) \\ 
137.382 &4(2,3)-3(2,2) & 69.2 & -4.56 & 18 & 18 (5) & -0.8 & -1.7/-0.8 & 0.8 & 1.2& 0.6& 0.15 (0.03) \\ 
171.688 &5(0,5)-4(0,4) & 24.7 & -4.14 & 14 & 69 (8) & -0.8 & -4.2/1.2 & 0.7 & 1.3& 0.7& 0.47 (0.10) \\ 
205.987 &6(0,6)-5(0,5) & 34.6 & -3.89 & 12 & 39 (3) & 0.3 & -5.4/2.6 & 1.1 & 1.5& 0.7& 0.40 (0.08) \\ 
206.158 &6(2,4)-5(2,3) & 87.3 & -3.95 & 12 & 21 (4) & -1.9 & -3.1/1.5 & 1.1 & 1.5& 0.7& 0.20 (0.04) \\ 
240.382 &7(2,6)-6(2,5) & 98.9 & -3.73 & 10 & 13 (3) & 0.7 & 0.7/1.6 & 1.0 & 1.6& 0.8& 0.12 (0.03) \\ 
240.549 &7(2,5)-6(2,4) & 98.9 & -3.73 & 10 & 18 (3) & -1.3 & -3.2/2.6 & 1.0 & 1.6& 0.8& 0.19 (0.04) \\ 
\hline 
 &&&&&\textbf{OCS ( $^1\Sigma^+$)}&&\\ 
 \hline 
85.139 &7-6 & 16.3 & -5.77 & 29 & 32 (1) & -0.1 & -7.0/4.0 & 1.4 & 1.2& 0.3& 0.26 (0.05) \\ 
97.301 &8-7 & 21.0 & -5.59 & 25 & 37 (1) & 0.2 & -5.8/3.8 & 1.2 & 1.2& 0.4& 0.31 (0.06) \\ 
109.463 &9-8 & 26.3 & -5.43 & 22 & 45 (1) & 0.5 & -6.0/3.7 & 1.1 & 1.2& 0.4& 0.35 (0.07) \\ 
133.785 &11-10 & 38.5 & -5.17 & 18 & 45 (4) & 0.8 & -5.3/2.6 & 0.9 & 1.2& 0.5& 0.35 (0.07) \\ 
145.946 &12-11 & 45.5 & -5.05 & 17 & 51 (3) & 0.2 & -6.2/3.4 & 0.8 & 1.3& 0.6& 0.45 (0.09) \\ 
158.107 &13-12 & 53.1 & -4.95 & 16 & 66 (5) & 1.1 & -4.8/2.6 & 0.7 & 1.3& 0.6& 0.52 (0.10) \\ 
170.267 &14-13 & 61.3 & -4.85 & 14 & 64 (6) & 1.9 & -5.0/3.3 & 0.7 & 1.3& 0.7& 0.54 (0.11) \\ 
206.745 &17-16 & 89.3 & -4.59 & 12 & 44 (2) & 1.5 & -5.3/2.6 & 1.1 & 1.5& 0.7& 0.45 (0.09) \\ 
218.903 &18-17 & 99.8 & -4.52 & 11 & 43 (2) & 0.5 & -4.9/4.7 & 1.1 & 1.5& 0.8& 0.43 (0.09) \\ 
231.060 &19-18 & 110.9 & -4.45 & 11 & 34 (3) & 0.6 & -4.5/2.6 & 1.0 & 1.6& 0.8& 0.32 (0.06) \\ 
243.218 &20-19 & 122.6 & -4.38 & 10 & 33 (3) & 0.7 & -5.1/2.6 & 1.0 & 1.6& 0.8& 0.33 (0.07) \\ 
255.374 &21-20 & 134.8 & -4.32 & 10 & 28 (3) & 0.8 & -3.8/2.6 & 0.9 & 1.6& 0.8& 0.29 (0.06) \\ 
267.530 &22-21 & 147.7 & -4.25 & 9 & 21 (5) & 0.8 & -1.5/0.8 & 2.2 & 1.7& 0.8& 0.30 (0.06) \\ 
291.839 &24-23 & 175.1 & -4.14 & 8 & 20 (4) & -1.1 & -3.2/0.9 & 2.0 & 1.9& 0.8& 0.28 (0.06) \\ 
\hline
\end{tabular}
\label{table:lines}
\end{table*}

\begin{table*}
\caption{Continued line properties of detected transitions}
\begin{tabular}{cccccccccccc}
\hline
 \textbf{Freq} & \textbf{Transition} & \boldmath$\mathrm{E_u}$  & \boldmath$\mathrm{log(A_{ij})}$ & \boldmath$\mathrm{\Theta_{MB}}$ & \boldmath$\mathrm{T_{A*,peak}}$ & \boldmath$\mathrm{V_{peak}}$ & \boldmath$\mathrm{V_{min}/V_{max}}$ &  \boldmath$\mathrm{\Delta V}$ & \boldmath$\mathrm{F_{eff}/B_{eff}}$ & \boldmath$\mathrm{\eta_{ff}}$ & \boldmath$\mathrm{\int T_{MB} dv}$ \\ 
 \textbf{GHz} & & \textbf{K} & & \textbf{\si{\arcsecond}} & \textbf{mK} &\textbf{km s$^{-1}$} & \textbf{km s$^{-1}$} & \textbf{km s$^{-1}$} & & & \textbf{K km s$^{-1}$} \\ \hline 
 &&&&&\textbf{SO ($^3\Sigma^-$)}&&\\ 
 \hline 
86.093 &2(2)-1(1) & 19.3 & -5.27 & 29 & 139 (1) & -0.1 & -20.5/4.0 & 1.4 & 1.2& 0.3& 1.07 (0.22) \\ 
99.299 &3(2)-2(1) & 9.2 & -4.94 & 25 & 1516 (2) & 0.2 & -21.0/5.0 & 1.2 & 1.2& 0.4& 11.31 (2.26) \\ 
100.029 &4(5)-4(4) & 38.6 & -5.96 & 25 & 14 (1) & 0.3 & -5.6/2.6 & 1.2 & 1.2& 0.4& 0.10 (0.02) \\ 
109.252 &2(3)-1(2) & 21.1 & -4.96 & 23 & 212 (2) & 0.5 & -15.6/3.7 & 1.1 & 1.2& 0.4& 1.64 (0.33) \\ 
129.138 &3(3)-2(2) & 25.5 & -4.64 & 19 & 315 (4) & 0.8 & -12.8/3.5 & 0.9 & 1.2& 0.5& 2.32 (0.46) \\ 
138.178 &4(3)-3(2) & 15.9 & -4.49 & 18 & 1873 (4) & 0.9 & -16.0/4.3 & 0.8 & 1.2& 0.6& 13.91 (2.78) \\ 
158.971 &3(4)-2(3) & 28.7 & -4.36 & 15 & 452 (5) & 1.1 & -12.8/4.1 & 0.7 & 1.3& 0.6& 3.67 (0.73) \\ 
172.181 &4(4)-3(3) & 33.8 & -4.23 & 14 & 389 (10) & 0.6 & -7.6/3.3 & 0.7 & 1.3& 0.7& 3.06 (0.61) \\ 
206.176 &4(5)-3(4) & 38.6 & -3.99 & 12 & 405 (4) & 0.3 & -14.4/3.7 & 1.1 & 1.5& 0.7& 3.97 (0.80) \\ 
215.220 &5(5)-4(4) & 44.1 & -3.91 & 11 & 404 (2) & 0.4 & -15.9/3.7 & 1.1 & 1.5& 0.8& 4.09 (0.82) \\ 
219.949 &6(5)-5(4) & 35.0 & -3.87 & 11 & 1293 (14) & 0.5 & -14.4/4.7 & 1.1 & 1.5& 0.8& 13.18 (2.64) \\ 
251.825 &5(6)-4(5) & 50.7 & -3.71 & 10 & 349 (23) & -0.2 & -5.8/2.6 & 0.9 & 1.6& 0.8& 4.01 (0.80) \\ 
258.255 &6(6)-5(5) & 56.5 & -3.67 & 10 & 246 (3) & -0.1 & -16.4/4.4 & 0.9 & 1.7& 0.8& 2.88 (0.58) \\ 
261.843 &7(6)-6(5) & 47.6 & -3.63 & 9 & 792 (4) & 0.8 & -16.2/4.4 & 0.9 & 1.7& 0.8& 8.30 (1.66) \\ 
286.340 &1(1)-1(0) & 15.2 & -4.84 & 9 & 35 (7) & 0.9 & -1.2/0.9 & 2.1 & 1.8& 0.8& 0.43 (0.09) \\ 
296.550 &6(7)-5(6) & 64.9 & -3.48 & 8 & 184 (3) & -1.0 & -19.2/3.0 & 2.0 & 1.9& 0.8& 2.69 (0.54) \\ 
301.286 &7(7)-6(6) & 71.0 & -3.46 & 8 & 105 (3) & -1.0 & -14.9/3.0 & 2.0 & 1.9& 0.9& 1.64 (0.33) \\ 
304.077 &8(7)-7(6) & 62.1 & -3.43 & 8 & 271 (6) & -0.9 & -14.8/3.0 & 2.0 & 2.0& 0.9& 4.27 (0.86) \\ 
309.502 &2(2)-2(1) & 19.3 & -4.84 & 8 & 31 (3) & 1.1 & -4.8/3.0 & 1.9 & 2.0& 0.9& 0.49 (0.10) \\ 
339.341 &3(3)-3(2) & 25.5 & -4.83 & 7 & 27 (4) & -0.5 & -4.1/1.2 & 1.8 & 2.3& 0.9& 0.45 (0.09) \\ 
340.714 &7(8)-6(7) & 81.2 & -3.29 & 7 & 122 (5) & -0.5 & -14.6/3.0 & 1.8 & 2.3& 0.9& 2.40 (0.48) \\ 
344.310 &8(8)-7(7) & 87.5 & -3.28 & 7 & 73 (8) & -0.5 & -5.7/1.3 & 1.7 & 2.4& 0.9& 1.23 (0.25) \\ 
346.528 &9(8)-8(7) & 78.8 & -3.26 & 7 & 181 (4) & -0.5 & -14.3/4.7 & 1.7 & 2.0& 0.9& 2.78 (0.56) \\ 
\hline 
 &&&&&\textbf{SO$_2$ ($^1A_1$)}&&\\ 
 \hline 
83.688 &8(1,7)-8(0,8) & 36.7 & -5.17 & 29 & 82 (1) & -1.6 & -14.2/2.6 & 1.4 & 1.0& 0.3& 0.54 (0.11) \\ 
104.029 &3(1,3)-2(0,2) & 7.7 & -5.00 & 24 & 157 (2) & -0.8 & -14.3/10.5 & 1.1 & 1.2& 0.4& 1.26 (0.25) \\ 
104.239 &10(1,9)-10(0,10) & 54.7 & -4.95 & 24 & 58 (1) & 0.3 & -7.5/2.6 & 1.1 & 1.2& 0.4& 0.47 (0.09) \\ 
131.014 &12(1,11)-12(0,12) & 76.4 & -4.73 & 19 & 37 (5) & -0.1 & -3.6/1.7 & 0.9 & 1.2& 0.5& 0.31 (0.06) \\ 
134.004 &8(2,6)-8(1,7) & 43.1 & -4.60 & 18 & 16 (3) & -0.9 & -4.4/0.9 & 1.7 & 1.2& 0.5& 0.14 (0.03) \\ 
135.696 &5(1,5)-4(0,4) & 15.7 & -4.66 & 18 & 245 (5) & 0.0 & -6.0/3.5 & 0.9 & 1.2& 0.6& 1.83 (0.37) \\ 
140.306 &6(2,4)-6(1,5) & 29.2 & -4.60 & 18 & 33 (4) & 0.1 & -3.2/1.8 & 0.8 & 1.2& 0.6& 0.24 (0.05) \\ 
151.378 &2(2,0)-2(1,1) & 12.6 & -4.73 & 16 & 27 (4) & 1.8 & -2.8/1.8 & 0.8 & 1.3& 0.6& 0.22 (0.04) \\ 
158.199 &3(2,2)-3(1,3) & 15.3 & -4.60 & 16 & 62 (5) & 0.4 & -4.0/1.9 & 0.7 & 1.3& 0.6& 0.41 (0.08) \\ 
160.827 &10(0,10)-9(1,9) & 49.7 & -4.40 & 15 & 251 (5) & 0.4 & -7.6/3.3 & 0.7 & 1.3& 0.6& 1.96 (0.39) \\ 
163.605 &14(1,13)-14(0,14) & 101.8 & -4.52 & 15 & 30 (6) & -1.0 & -1.7/1.2 & 0.7 & 1.3& 0.6& 0.20 (0.04) \\ 
165.144 &5(2,4)-5(1,5) & 23.6 & -4.51 & 15 & 59 (11) & 0.5 & -2.4/1.9 & 0.7 & 1.3& 0.6& 0.47 (0.10) \\ 
165.225 &7(1,7)-6(0,6) & 27.1 & -4.38 & 15 & 278 (6) & 0.5 & -7.3/3.3 & 0.7 & 1.3& 0.6& 2.25 (0.45) \\ 
200.809 &16(1,15)-16(0,16) & 130.7 & -4.33 & 12 & 11 (2) & -0.9 & -0.9/0.3 & 1.2 & 1.5& 0.7& 0.07 (0.01) \\ 
203.391 &12(0,12)-11(1,11) & 70.1 & -4.06 & 12 & 147 (3) & 0.3 & -12.4/14.1 & 1.1 & 1.5& 0.7& 1.50 (0.30) \\ 
205.300 &11(2,10)-11(1,11) & 70.2 & -4.27 & 12 & 21 (3) & -0.8 & -3.1/2.6 & 1.1 & 1.5& 0.7& 0.21 (0.04) \\ 
208.700 &3(2,2)-2(1,1) & 15.3 & -4.17 & 12 & 90 (3) & 0.4 & -6.4/3.7 & 1.1 & 1.5& 0.7& 0.81 (0.16) \\ 
221.965 &11(1,11)-10(0,10) & 60.4 & -3.94 & 11 & 150 (2) & -0.6 & -13.2/3.7 & 1.1 & 1.5& 0.8& 1.60 (0.32) \\ 
225.153 &13(2,12)-13(1,13) & 93.0 & -4.19 & 11 & 13 (2) & -0.5 & -3.6/1.6 & 1.0 & 1.5& 0.8& 0.14 (0.03) \\ 
235.151 &4(2,2)-3(1,3) & 19.0 & -4.11 & 10 & 83 (3) & -0.4 & -6.4/2.6 & 1.0 & 1.6& 0.8& 0.82 (0.16) \\ 
\hline
\end{tabular}
\label{table:lines2}
\end{table*}

\begin{table*}
\caption{Continued line properties of detected transitions}
\begin{tabular}{cccccccccccc}
\hline
 \textbf{Freq} & \textbf{Transition} & \boldmath$\mathrm{E_u}$  & \boldmath$\mathrm{log(A_{ij})}$ & \boldmath$\mathrm{\Theta_{MB}}$ & \boldmath$\mathrm{T_{A*,peak}}$ & \boldmath$\mathrm{V_{peak}}$ & \boldmath$\mathrm{V_{min}/V_{max}}$ &  \boldmath$\mathrm{\Delta V}$ & \boldmath$\mathrm{F_{eff}/B_{eff}}$ & \boldmath$\mathrm{\eta_{ff}}$ & \boldmath$\mathrm{\int T_{MB} dv}$ \\ 
 \textbf{GHz} & & \textbf{K} & & \textbf{\si{\arcsecond}} & \textbf{mK} &\textbf{km s$^{-1}$} & \textbf{km s$^{-1}$} & \textbf{km s$^{-1}$} & & & \textbf{K km s$^{-1}$} \\ \hline 
  &&&&&\textbf{SO$_2$ ($^1A_1$)}&&\\ 
 \hline
 236.216 &16(1,15)-15(2,14) & 130.7 & -4.12 & 10 & 13 (3) & -1.4 & -2.4/-1.4 & 1.0 & 1.6& 0.8& 0.11 (0.02) \\ 
241.615 &5(2,4)-4(1,3) & 23.6 & -4.07 & 10 & 89 (3) & -0.3 & -7.1/2.6 & 1.0 & 1.6& 0.8& 0.88 (0.18) \\ 
254.280 &6(3,3)-6(2,4) & 41.4 & -3.94 & 10 & 23 (3) & 0.8 & -3.9/1.7 & 0.9 & 1.6& 0.8& 0.23 (0.05) \\ 
255.553 &4(3,1)-4(2,2) & 31.3 & -4.03 & 10 & 22 (4) & -1.1 & -2.9/0.8 & 0.9 & 1.6& 0.8& 0.24 (0.05) \\ 
256.246 &5(3,3)-5(2,4) & 35.9 & -3.97 & 10 & 21 (4) & -1.1 & -2.0/1.7 & 0.9 & 1.6& 0.8& 0.21 (0.04) \\ 
257.099 &7(3,5)-7(2,6) & 47.8 & -3.91 & 10 & 17 (3) & 0.8 & -3.8/2.6 & 0.9 & 1.7& 0.8& 0.22 (0.04) \\ 
271.529 &7(2,6)-6(1,5) & 35.5 & -3.96 & 9 & 47 (3) & -1.4 & -8.0/3.0 & 2.2 & 1.7& 0.8& 0.67 (0.13) \\ 
282.036 &6(2,4)-5(1,5) & 29.2 & -4.00 & 9 & 56 (3) & -1.2 & -5.5/3.0 & 2.1 & 1.8& 0.8& 0.74 (0.15) \\ 
283.464 &16(0,16)-15(1,15) & 121.0 & -3.57 & 9 & 29 (4) & -1.2 & -3.4/0.9 & 2.1 & 1.8& 0.8& 0.41 (0.08) \\ 
298.576 &9(2,8)-8(1,7) & 51.0 & -3.84 & 8 & 24 (3) & -1.0 & -3.0/3.0 & 2.0 & 1.9& 0.8& 0.34 (0.07) \\ 
313.279 &3(3,1)-2(2,0) & 27.6 & -3.47 & 8 & 14 (3) & 1.1 & -0.8/1.1 & 1.9 & 2.1& 0.9& 0.22 (0.04) \\ 
334.673 &8(2,6)-7(1,7) & 43.1 & -3.90 & 7 & 37 (4) & -0.6 & -4.2/1.2 & 1.8 & 2.3& 0.9& 0.49 (0.10) \\ 
\hline
 &&&&&\textbf{$^{34}$SO ($^3\Sigma^-$)}&&\\ 
 \hline 
84.411 &2(2)-1(1) & 19.2 & -5.30 & 29 & 5 (1) & -0.2 & -1.6/1.2 & 1.4 & 1.0& 0.3& 0.03 (0.01) \\ 
97.715 &3(2)-2(1) & 9.1 & -4.96 & 25 & 75 (2) & 0.2 & -8.2/3.8 & 1.2 & 1.2& 0.4& 0.56 (0.11) \\ 
106.743 &2(3)-1(2) & 20.9 & -4.99 & 23 & 7 (2) & -1.8 & -2.9/-0.7 & 1.1 & 1.2& 0.4& 0.05 (0.01) \\ 
135.775 &4(3)-3(2) & 15.6 & -4.51 & 18 & 87 (4) & 0.9 & -6.9/3.5 & 0.9 & 1.2& 0.6& 0.56 (0.11) \\ 
155.506 &3(4)-2(3) & 28.4 & -4.39 & 16 & 25 (6) & 1.1 & -1.2/1.1 & 0.8 & 1.3& 0.6& 0.16 (0.03) \\ 
201.846 &4(5)-3(4) & 38.1 & -4.02 & 12 & 15 (3) & -0.9 & -3.2/2.6 & 1.2 & 1.5& 0.7& 0.19 (0.04) \\ 
211.013 &5(5)-4(4) & 43.5 & -3.94 & 12 & 14 (3) & 1.5 & -1.8/2.6 & 1.1 & 1.5& 0.8& 0.14 (0.03) \\ 
215.839 &6(5)-5(4) & 34.4 & -3.89 & 11 & 49 (3) & 0.4 & -6.1/3.7 & 1.1 & 1.5& 0.8& 0.47 (0.09) \\ 
\hline 
 &&&&&\textbf{$^{34}$SO$_2$ ($^1A_1$)}&&\\ 
 \hline 
133.471 &5(1,5)-4(0,4) & 15.6 & -4.68 & 18 & 12 (3) & 1.7 & 0.8/1.7 & 0.9 & 1.2& 0.5& 0.07 (0.01) \\ 
162.775 &7(1,7)-6(0,6) & 27.0 & -4.40 & 15 & 24 (6) & -1.0 & -1.0/2.6 & 0.7 & 1.3& 0.6& 0.19 (0.04) \\ 
\hline
\end{tabular}
\label{table:lines3}
\end{table*}

\section{Detected Lines}
\label{appendix:profiles}
\begin{figure*}
\includegraphics[width=0.9\textwidth]{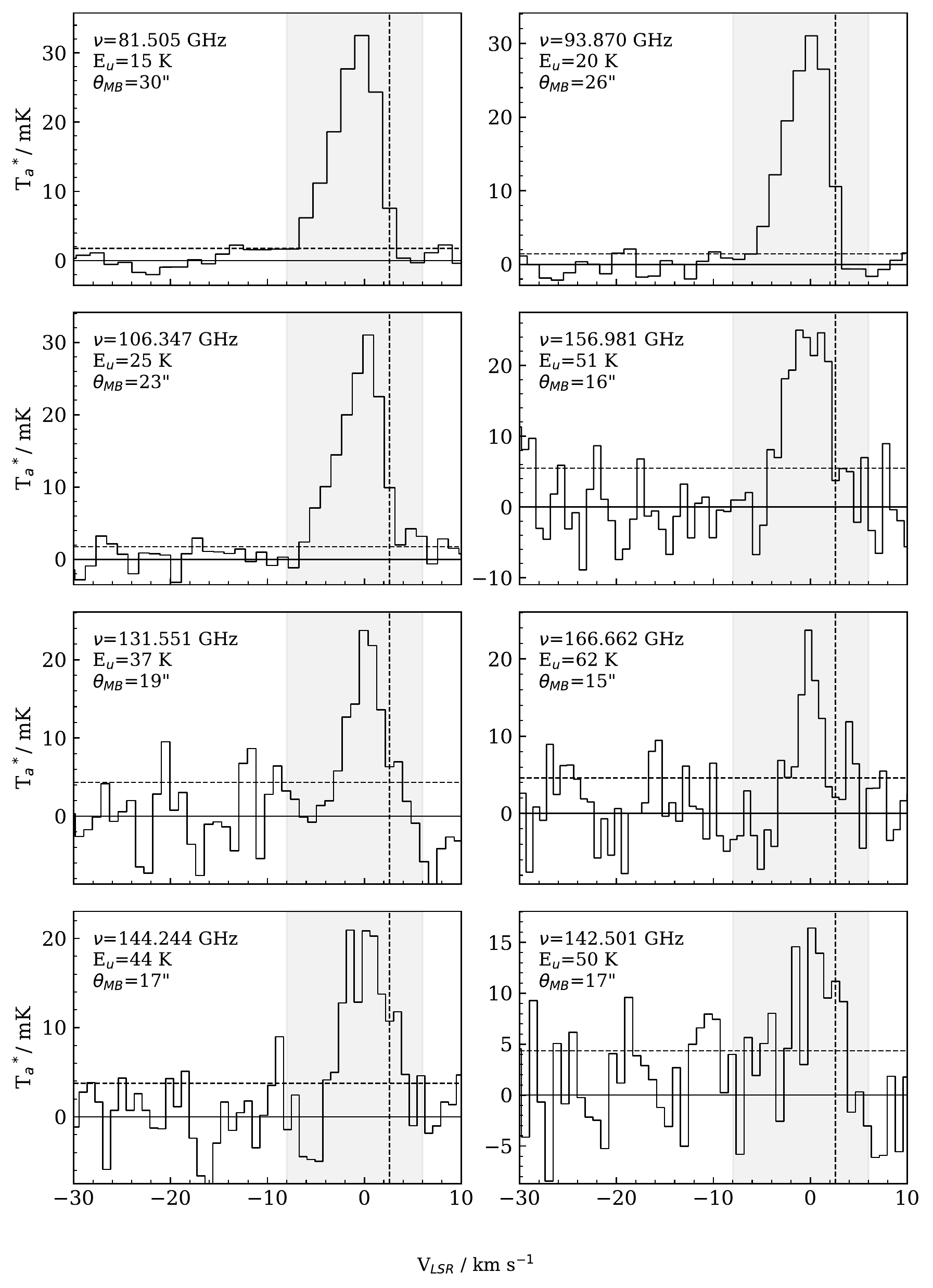}
\caption{Detected CCS lines. Rows share Y-axis values which are given in antenna temperature. The vertical black dashed line indicates the location of the local standard of rest velocity \SI{2.6}{\kilo\metre\per\second}. The horizontal dashed line shows the 1$\sigma$ level for that spectrum. The shaded area shows the moderate velocity regime over which the spectra were integrated.\label{fig:ccsspectra}}
\end{figure*}

\begin{figure*}
\includegraphics[width=0.9\textwidth]{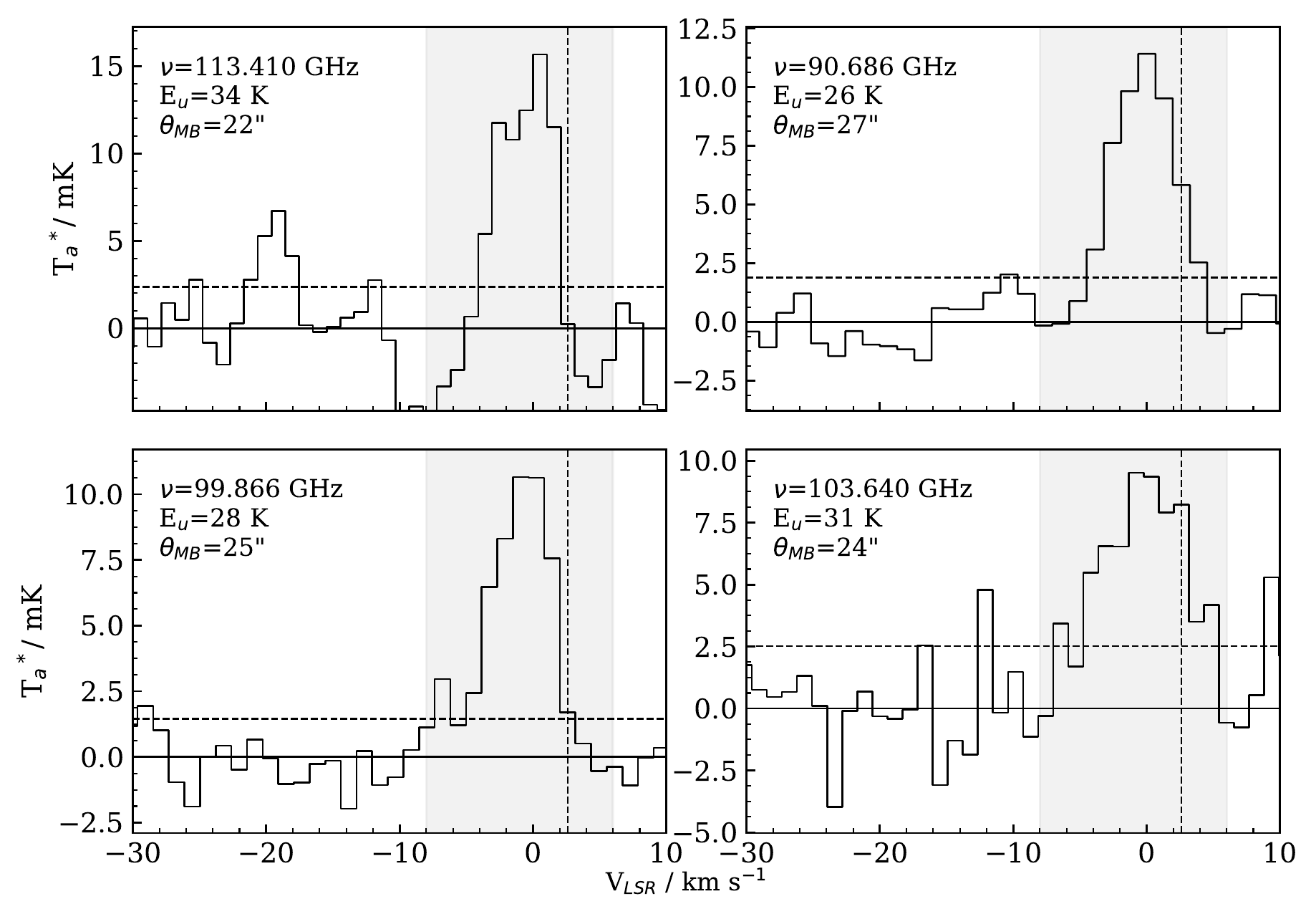}
\label{fig:ccsspectra2}
\caption{Detected CCS lines cont.}
\end{figure*}

\begin{figure*}
\includegraphics[width=0.9\textwidth]{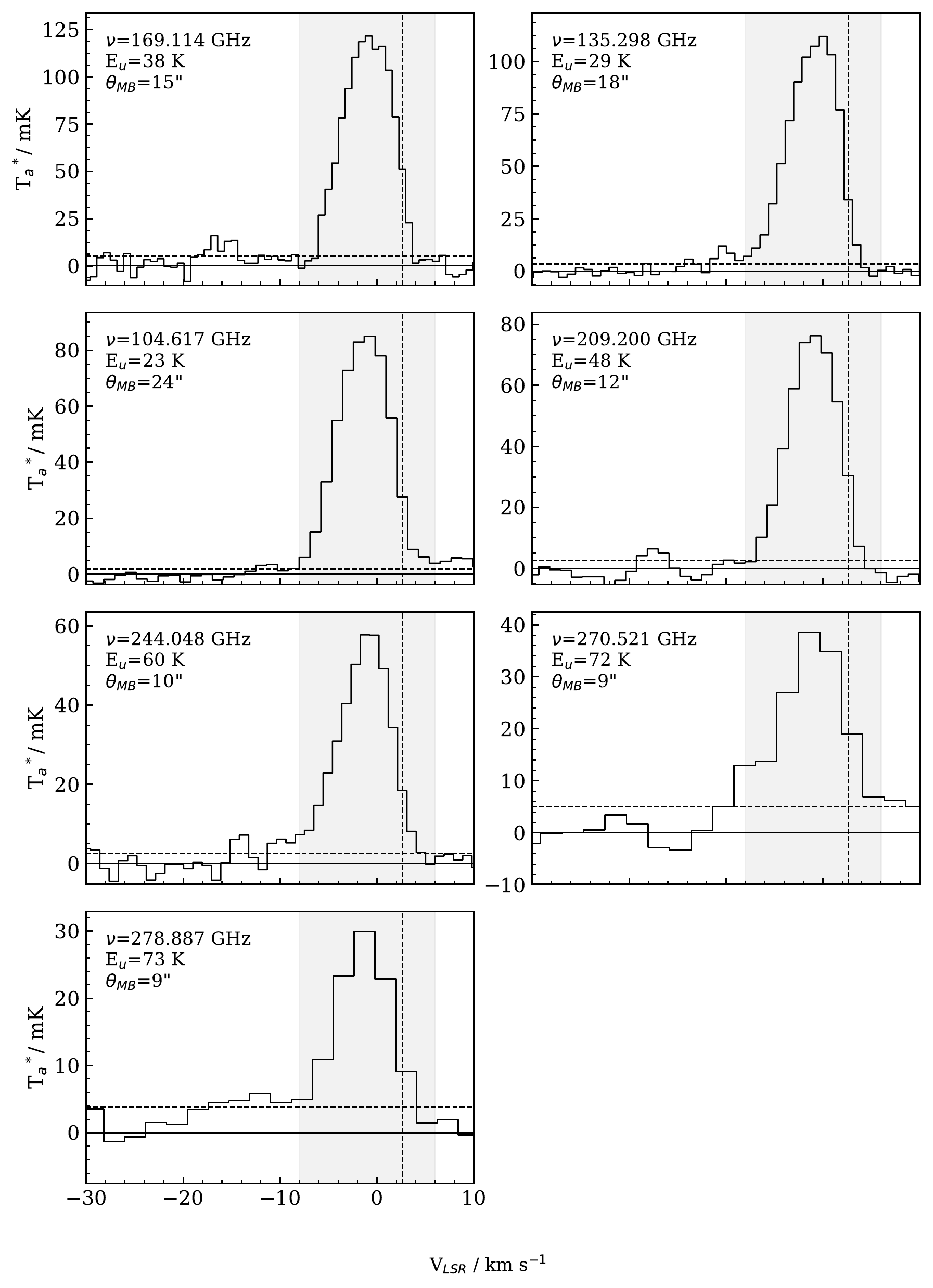}
\label{fig:oh2csspectra1}
\caption{Similar to Figure~\ref{fig:ccsspectra} for detected ortho H$_2$CS lines.}
\end{figure*}

\begin{figure*}
\includegraphics[width=0.9\textwidth]{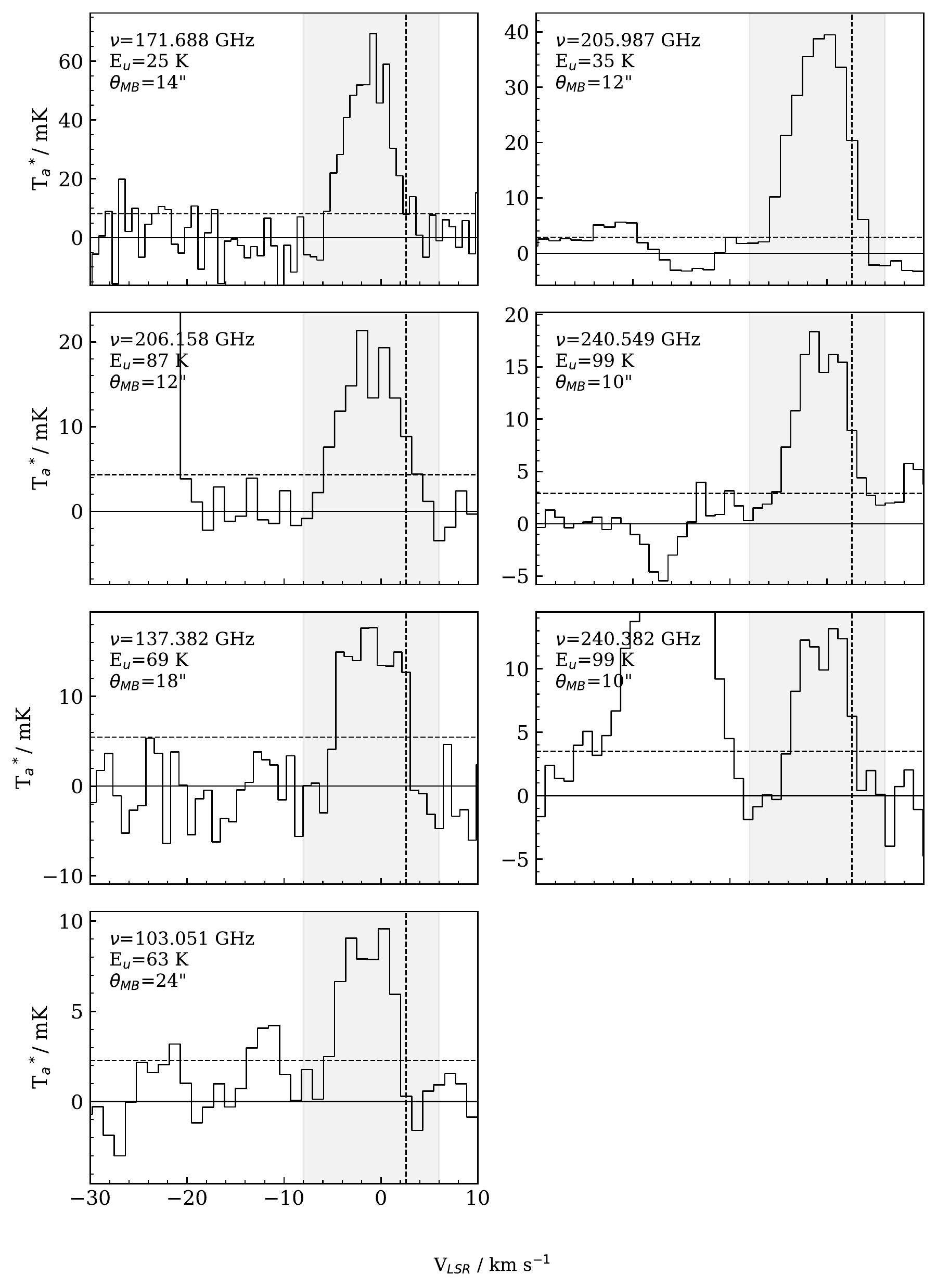}
\label{fig:ph2csspectra1}
\caption{Similar to Figure~\ref{fig:ccsspectra} for detected para H$_2$CS lines.}
\end{figure*}

\begin{figure*}
\includegraphics[width=0.9\textwidth]{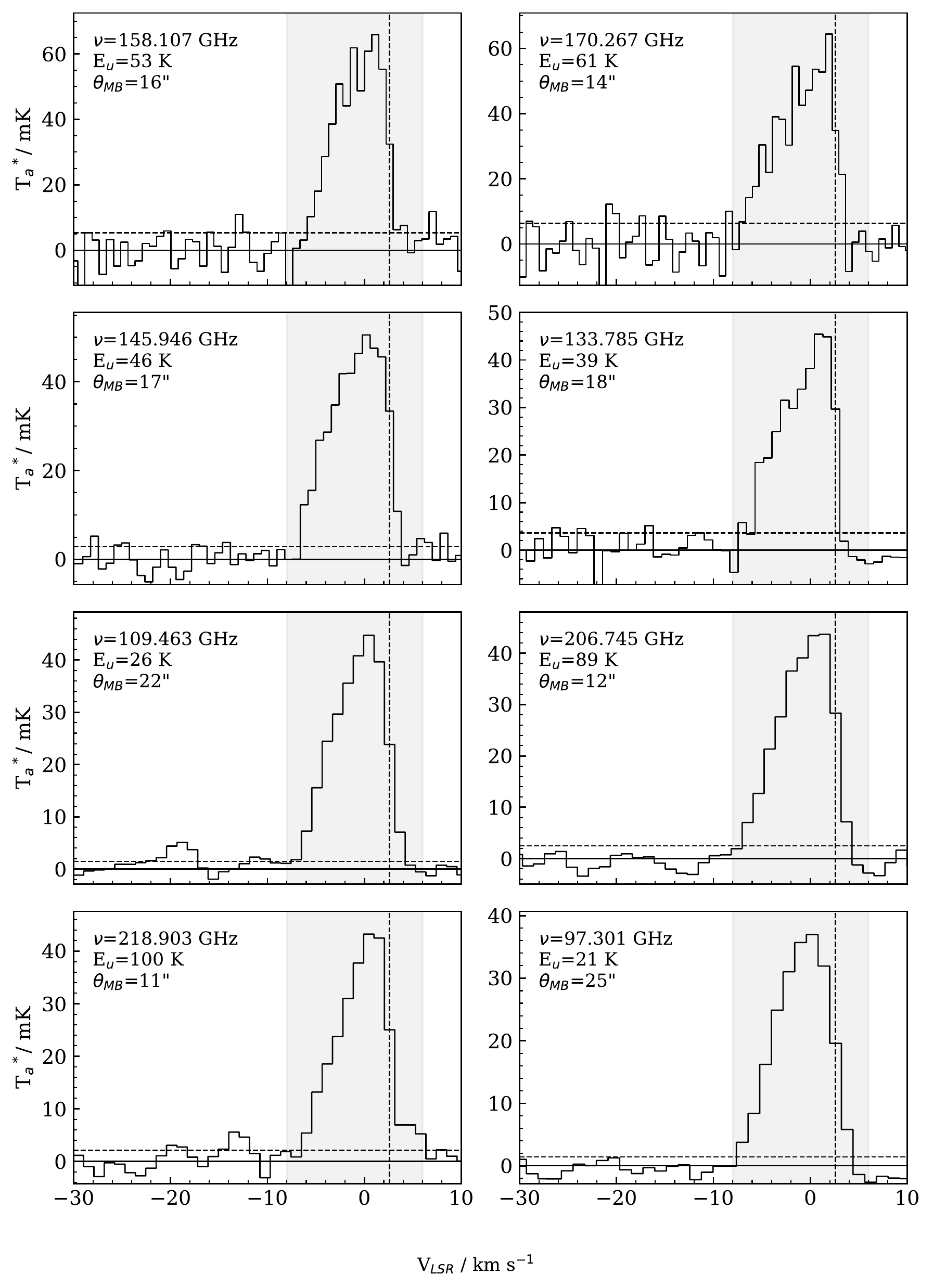}
\label{fig:ocsspectra}
\caption{Similar to Figure~\ref{fig:ccsspectra} for detected OCS lines.}
\end{figure*}

\begin{figure*}
\includegraphics[width=0.9\textwidth]{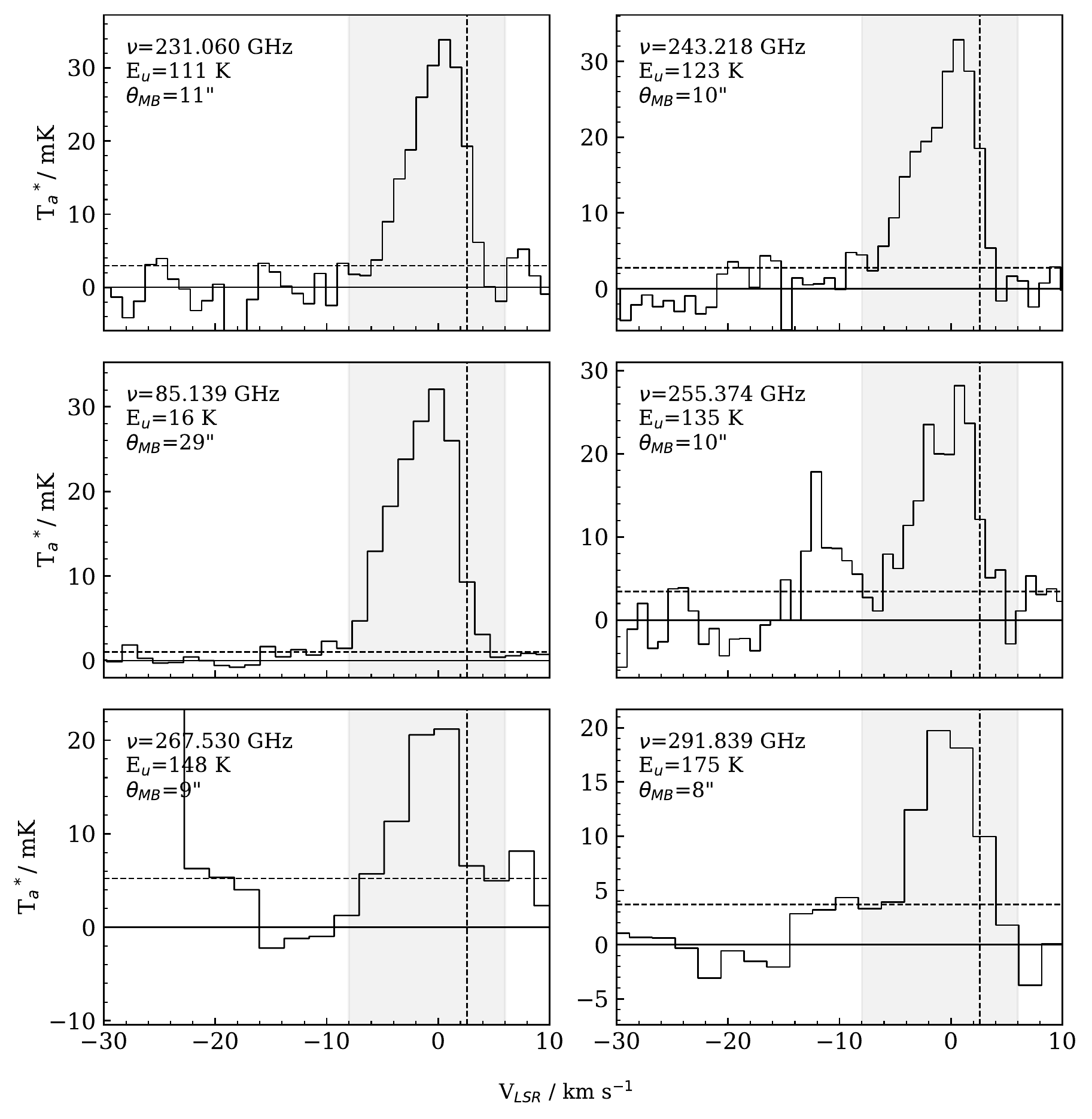}
\label{fig:ocsspectra2}
\caption{Detected OCS lines cont.}
\end{figure*}

\begin{figure*}
\includegraphics[width=0.9\textwidth]{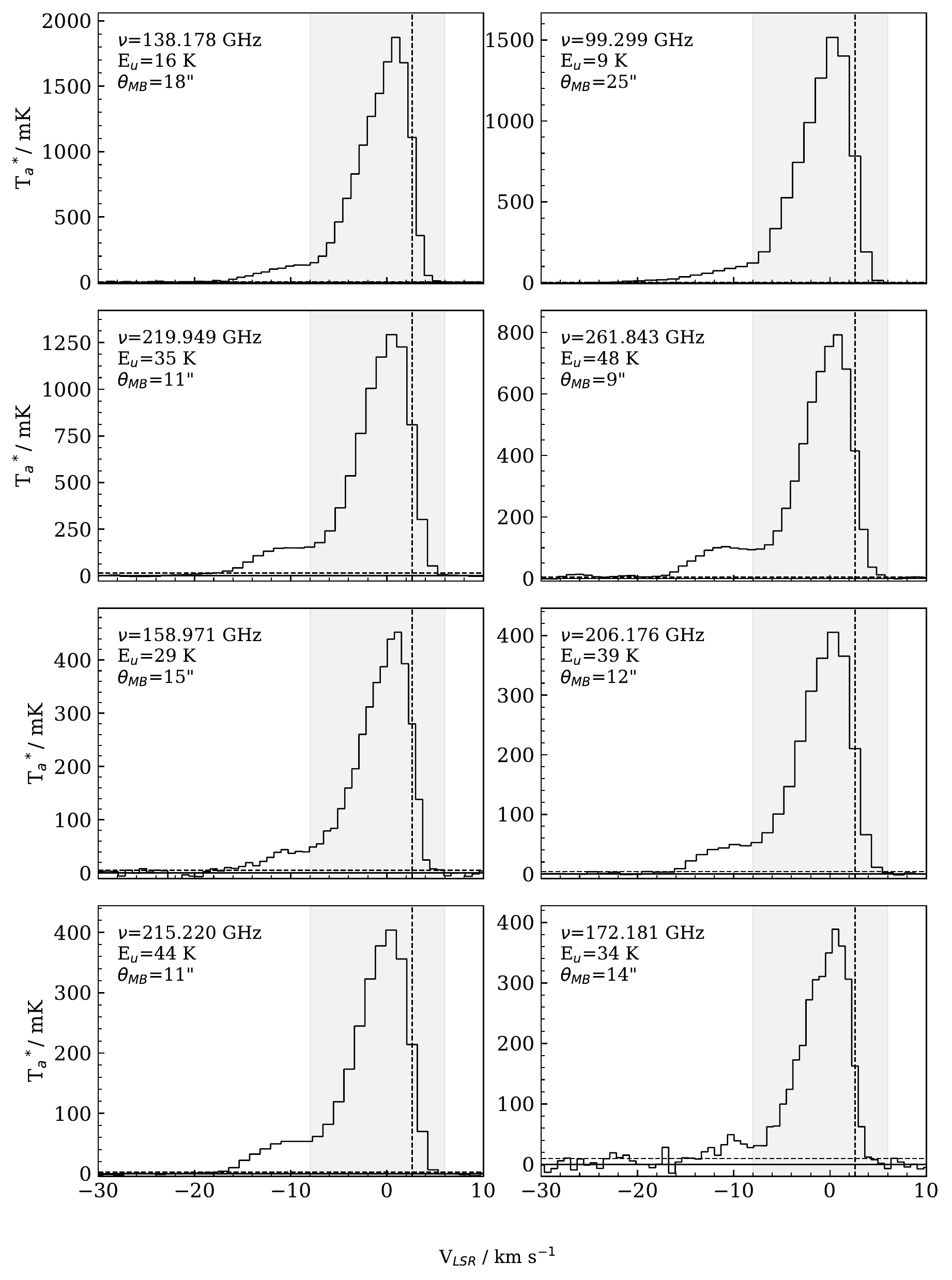}
\label{fig:sospectra}
\caption{Similar to Figure~\ref{fig:ccsspectra} for detected SO lines. The SO spectra often show a second peak more blue shifted than the peak traced by the grey histogram. These are analysed in Section~\ref{sec:sosecondary}.}
\end{figure*}

\begin{figure*}
\includegraphics[width=0.9\textwidth]{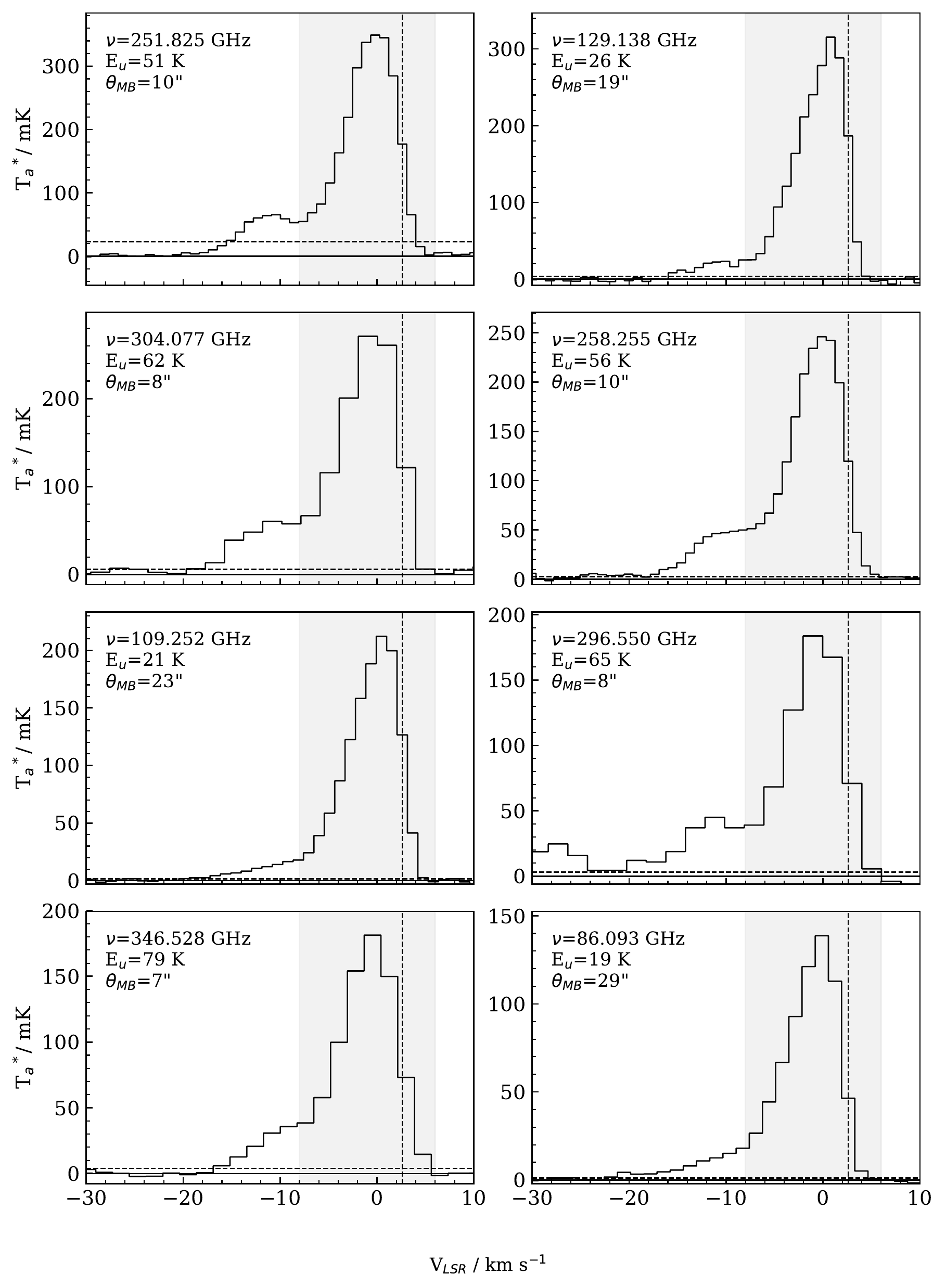}
\label{fig:sospectra2}
\caption{Detected SO lines cont.}
\end{figure*}

\begin{figure*}
\includegraphics[width=0.9\textwidth]{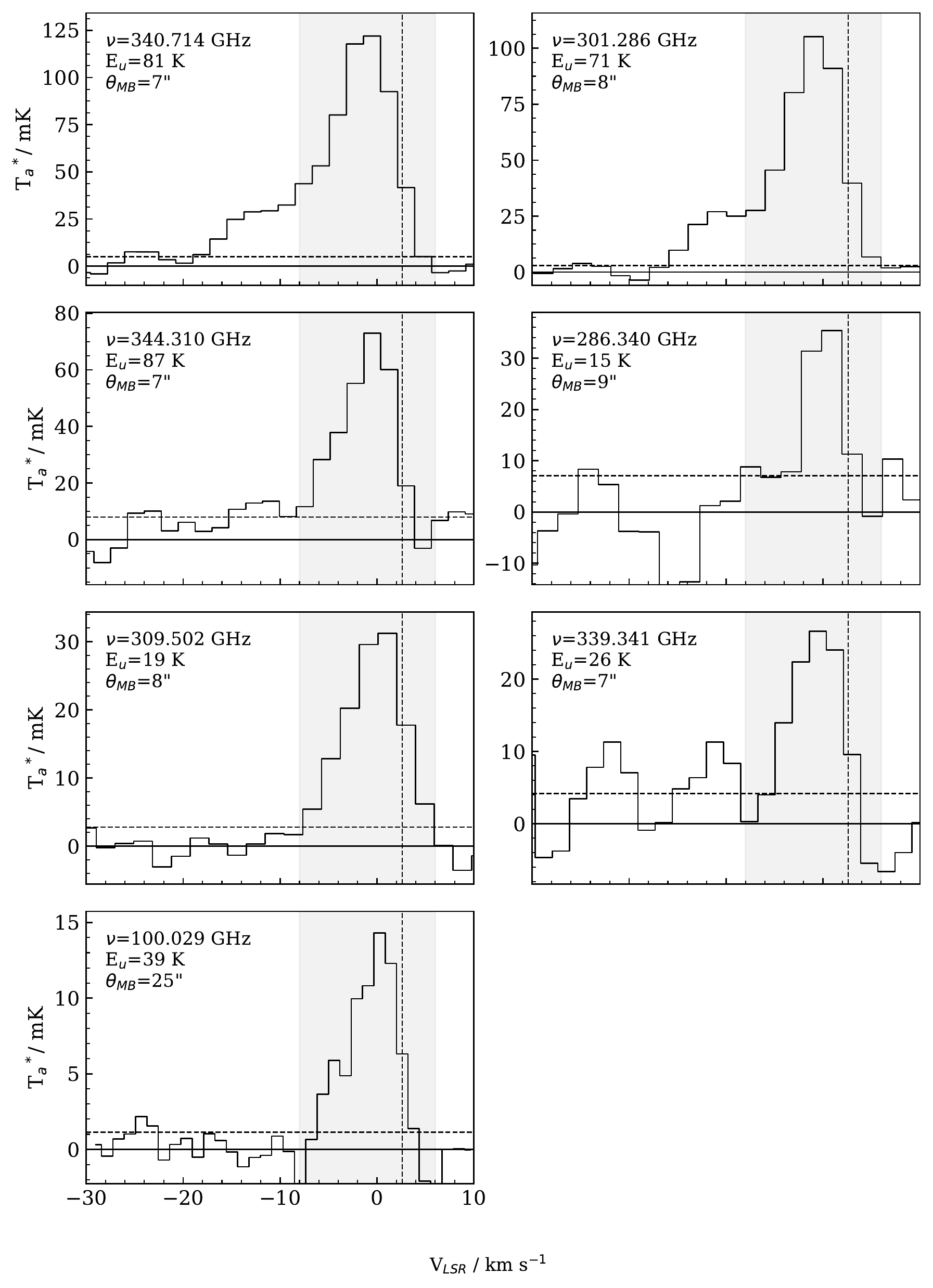}
\label{fig:sospectra3}
\caption{Detected SO lines cont.}
\end{figure*}

\begin{figure*}
\includegraphics[width=0.9\textwidth]{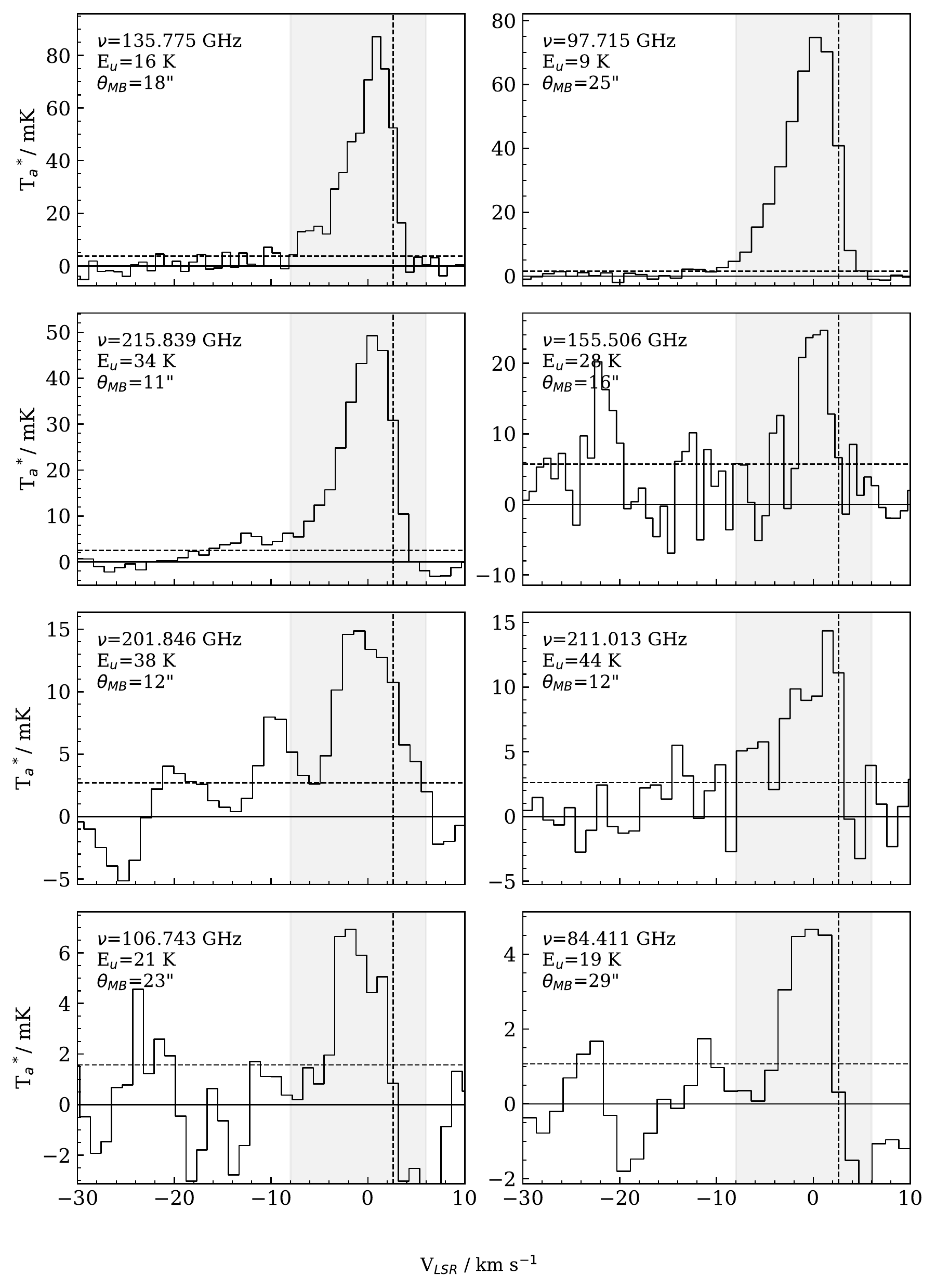}
\label{fig:34sospectra}
\caption{Similar to Figure~\ref{fig:ccsspectra} for detected $^{34}$SO lines.}
\end{figure*}

\begin{figure*}
\includegraphics[width=0.9\textwidth]{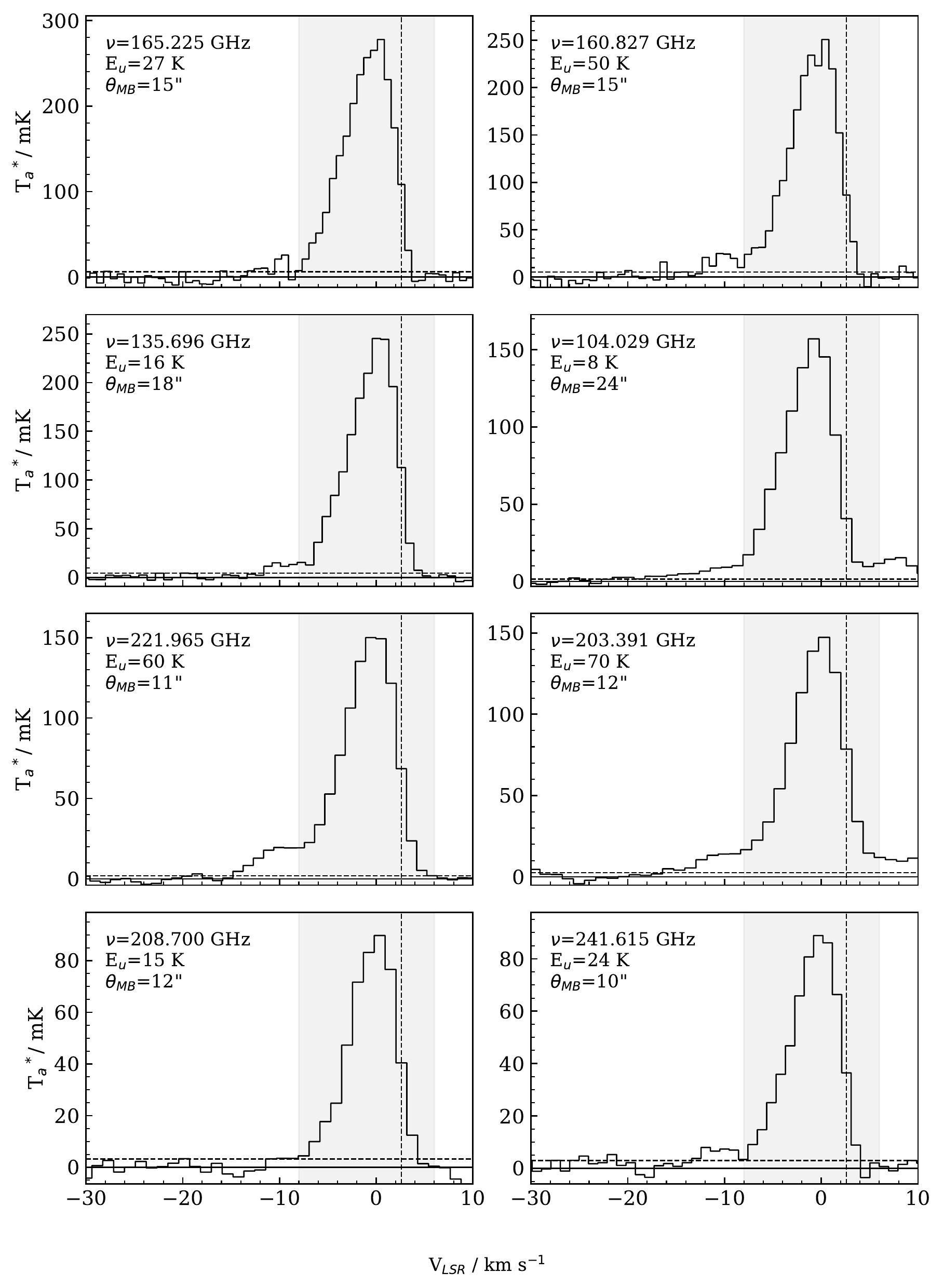}
\label{fig:so2spectra}
\caption{Similar to Figure~\ref{fig:ccsspectra} for detected SO$_2$ lines. A small number of lines show a secondary peak similar to that seen for SO and are analysed in Section~\ref{sec:sosecondary}.}
\end{figure*}

\begin{figure*}
\includegraphics[width=0.9\textwidth]{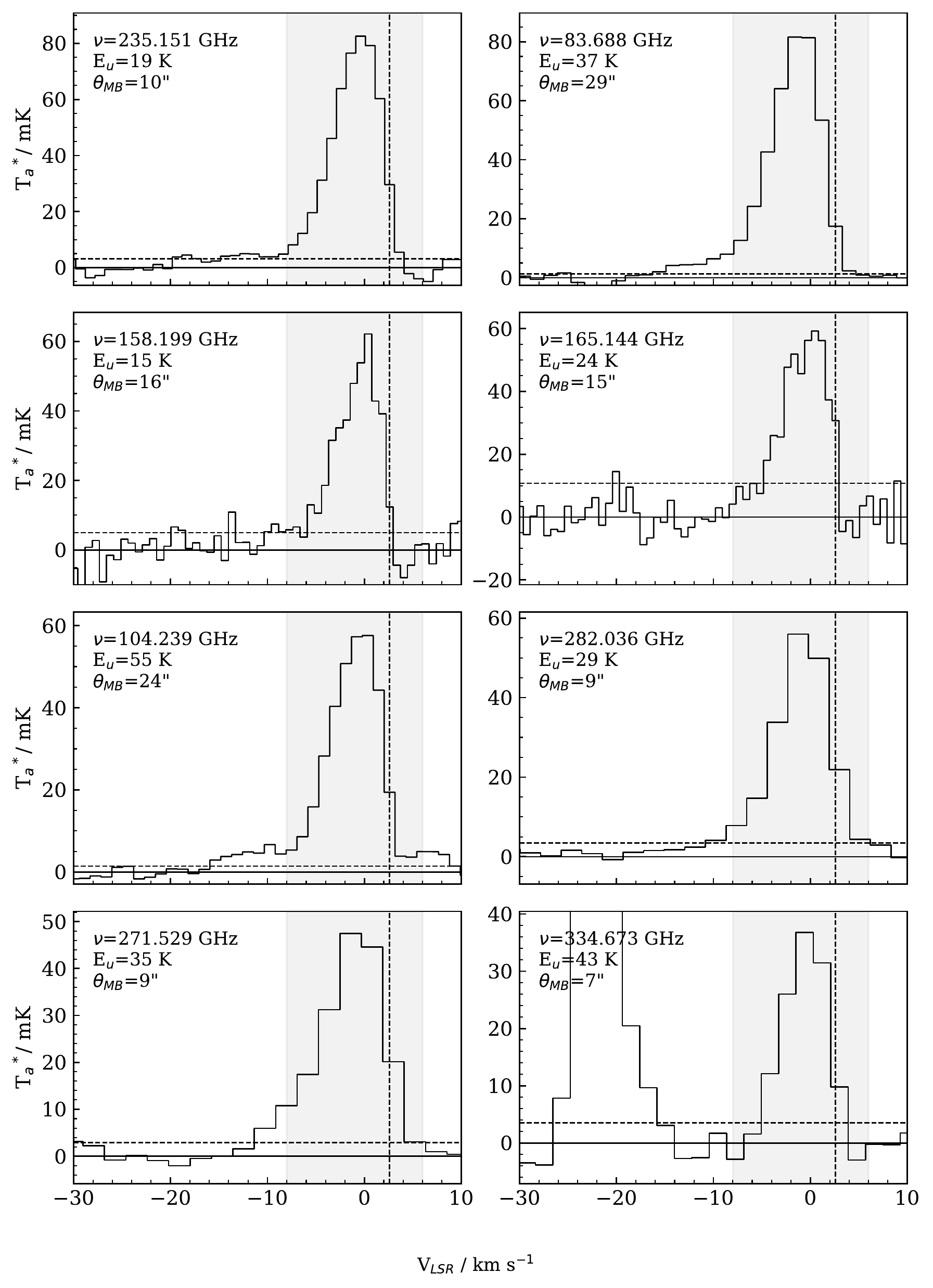}
\label{fig:so2spectra2}
\caption{ Detected SO$_2$ lines cont.}
\end{figure*}

\begin{figure*}
\includegraphics[width=0.9\textwidth]{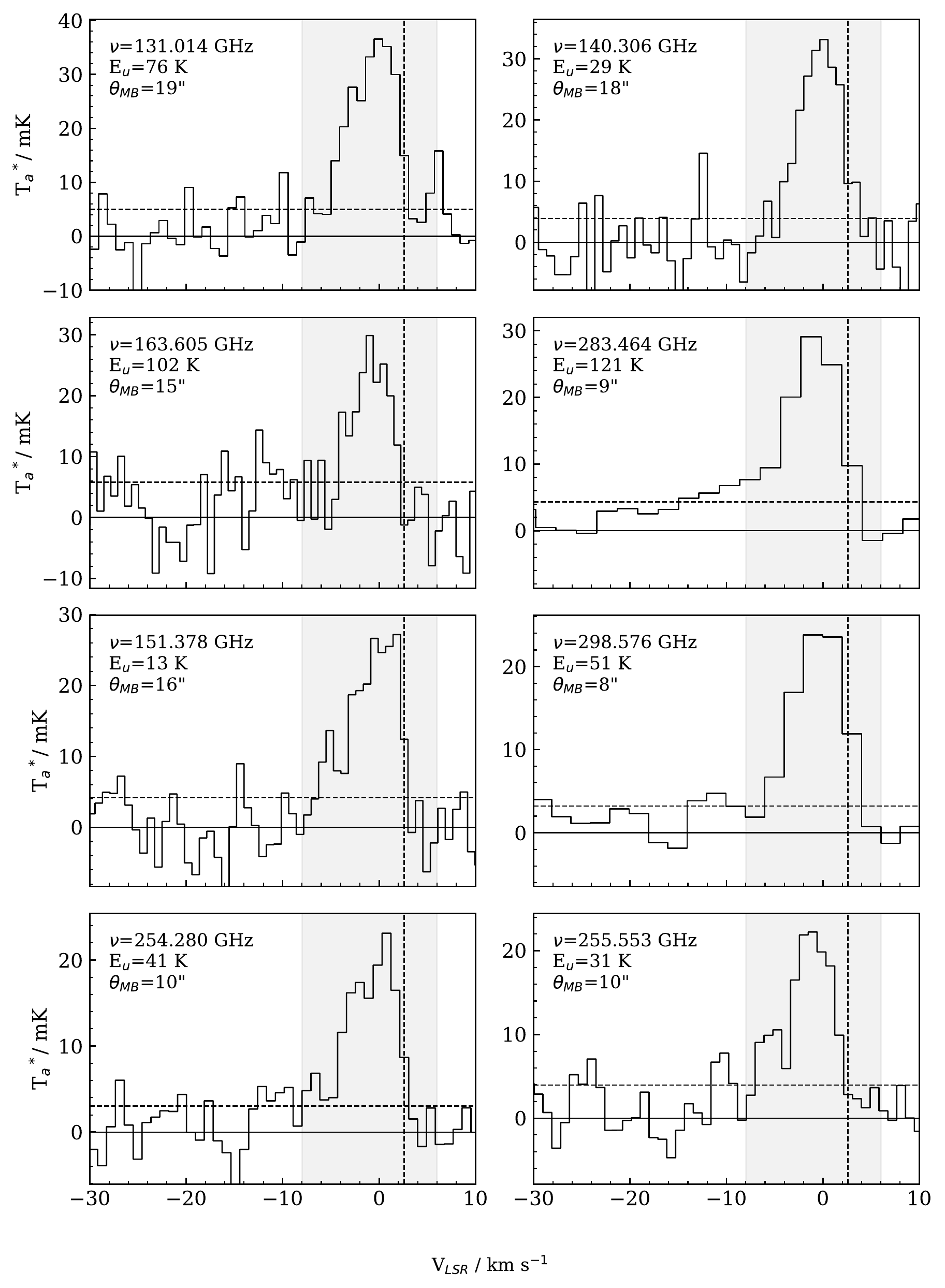}
\label{fig:so2spectra3}
\caption{Detected SO$_2$ lines cont.}
\end{figure*}

\begin{figure*}
\includegraphics[width=0.9\textwidth]{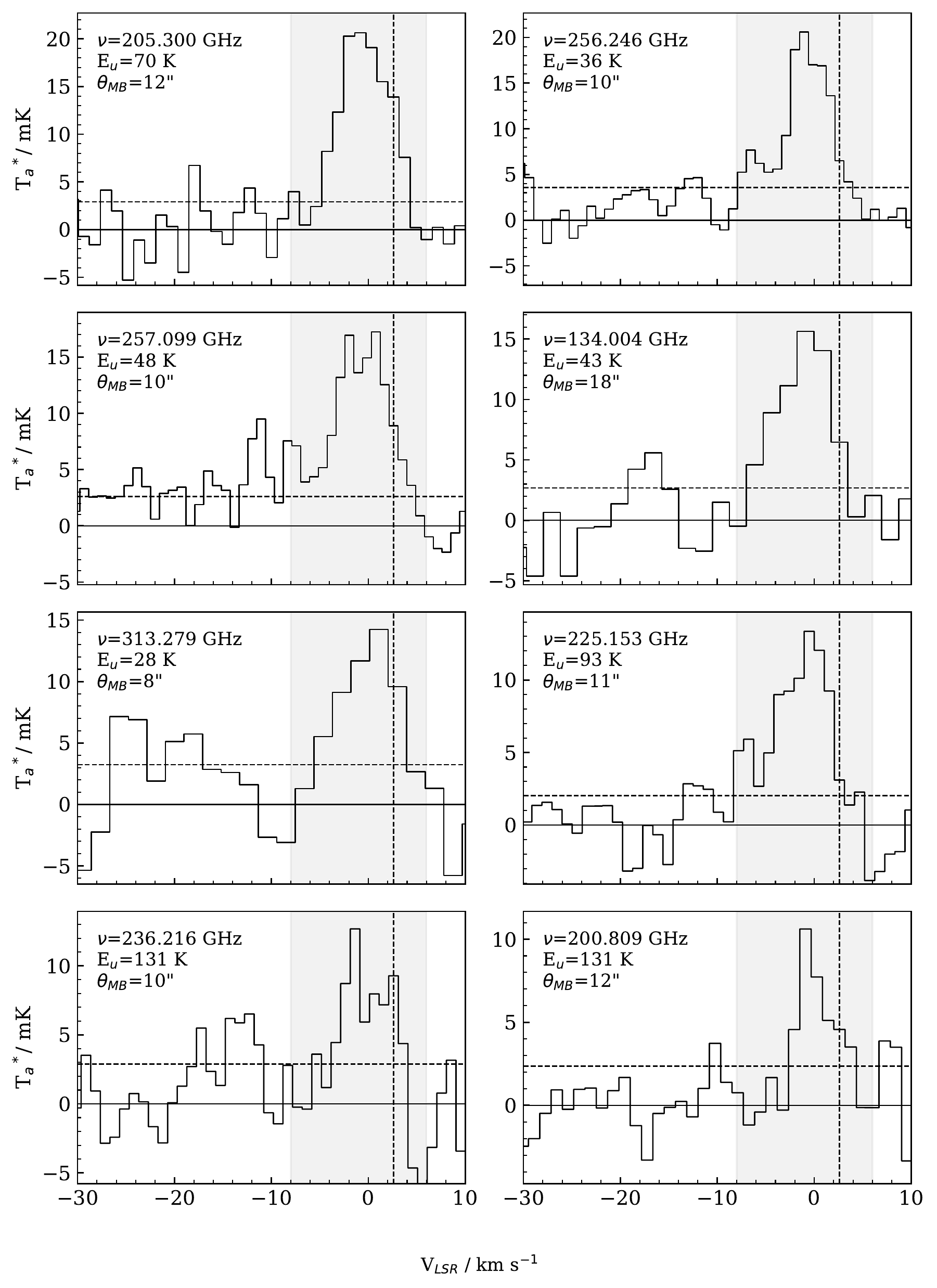}
\label{fig:so2spectra4}
\caption{Detected SO$_2$ lines cont.}
\end{figure*}

\begin{figure*}
\includegraphics[width=0.9\textwidth]{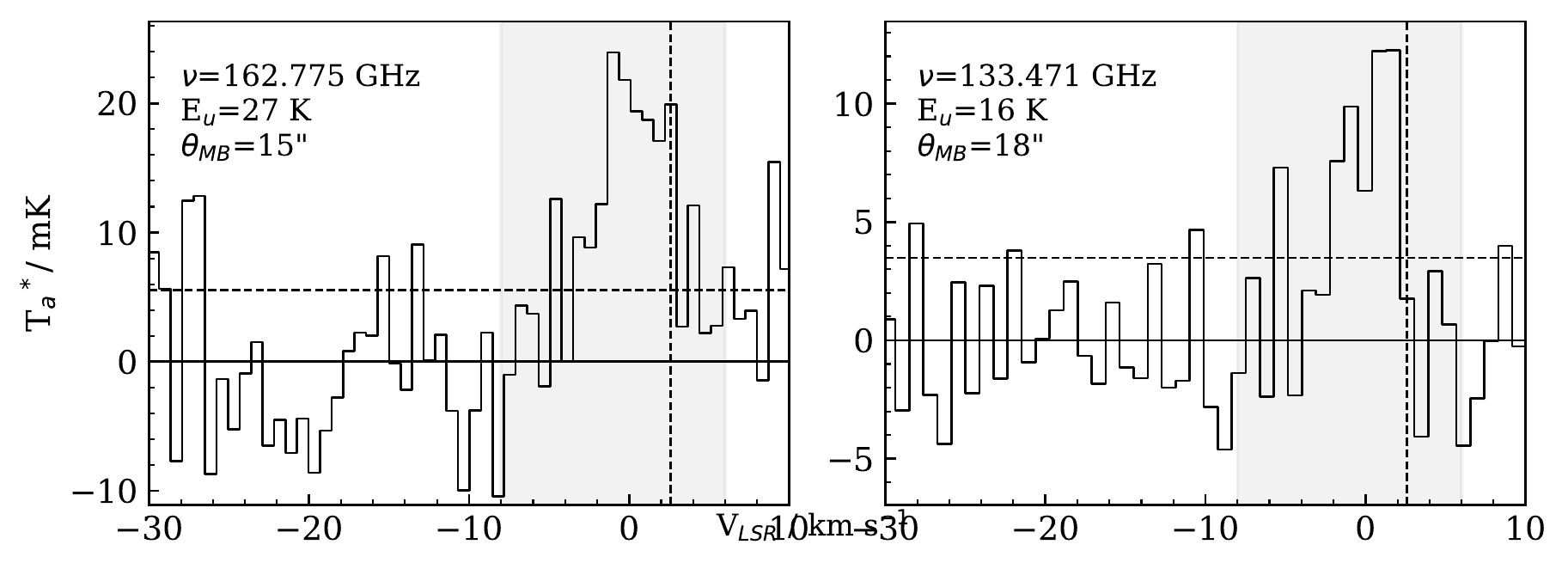}
\caption{Similar to Figure~\ref{fig:ccsspectra} for detected $^{34}$SO$_2$ lines.\label{fig:34so2spectra}}
\end{figure*}

The spectra used for this work are shown in Figures~\ref{fig:ccsspectra} to \ref{fig:34so2spectra} labelled with their frequency, upper state energy and the IRAM-30m beam size at that frequency. Spectra are organized by species and intensity.


\section{RADEX Probability Distributions}
\label{appendix:mcmc}
\begin{figure*}
\includegraphics[width=1.0\textwidth]{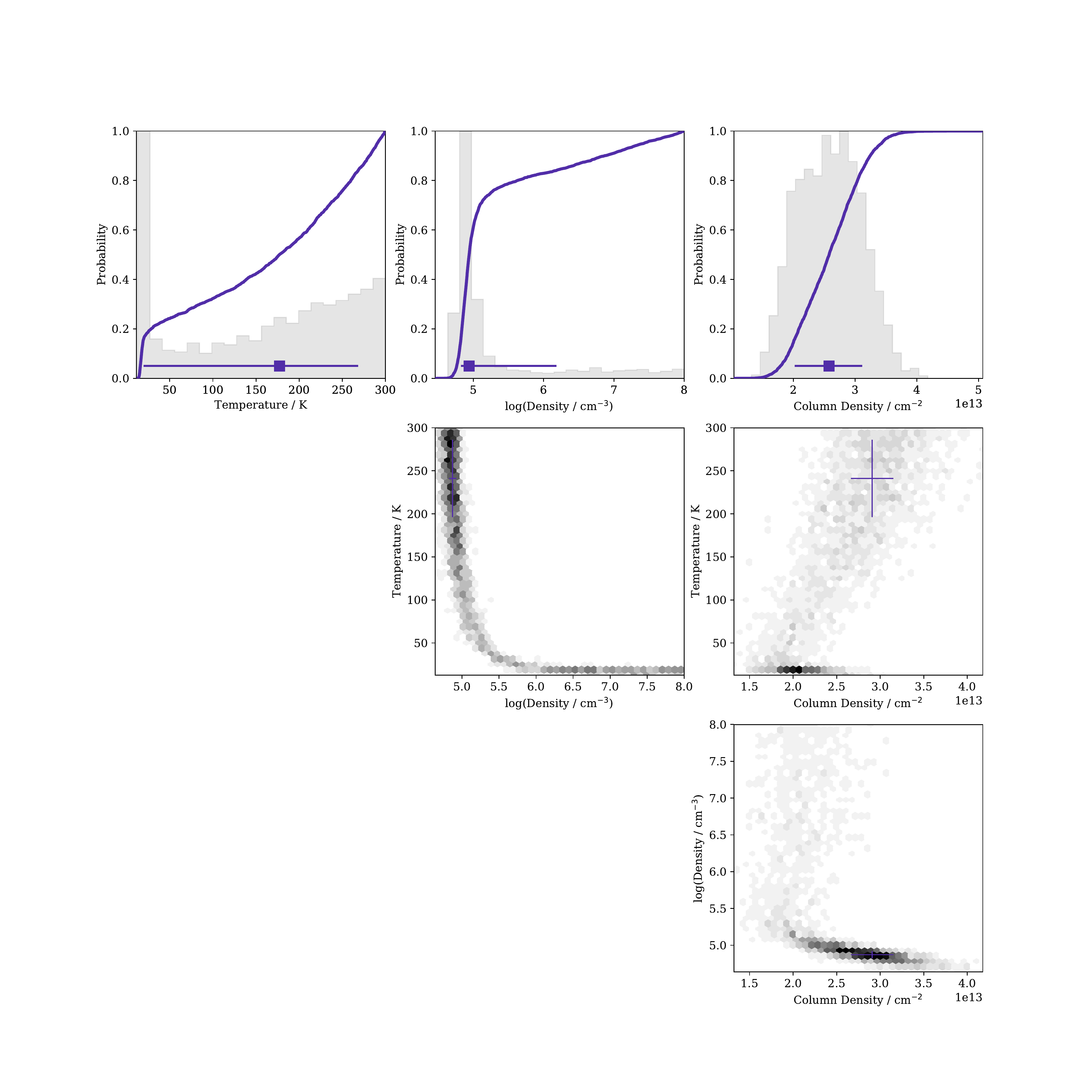}
\caption{Probability distributions for o-H$_2$CS. The horizontal spans in the top row and the crosses in the lower panels show the position of the median of the distribution and 67\% probability intervals for each parameter. The grey histograms show the probability density and the line shows the cumulative probability distribution. The grey scale in the lower panels shows the joint probablity distribution of each parameter pair with darker areas representing higher likelihoods. The column densities demonstrate the general trend of having a most likely value that is not strongly dependent on density and temperature. \label{fig:h2cschain}}
\end{figure*}

\begin{figure*}
\includegraphics[width=1.0\textwidth]{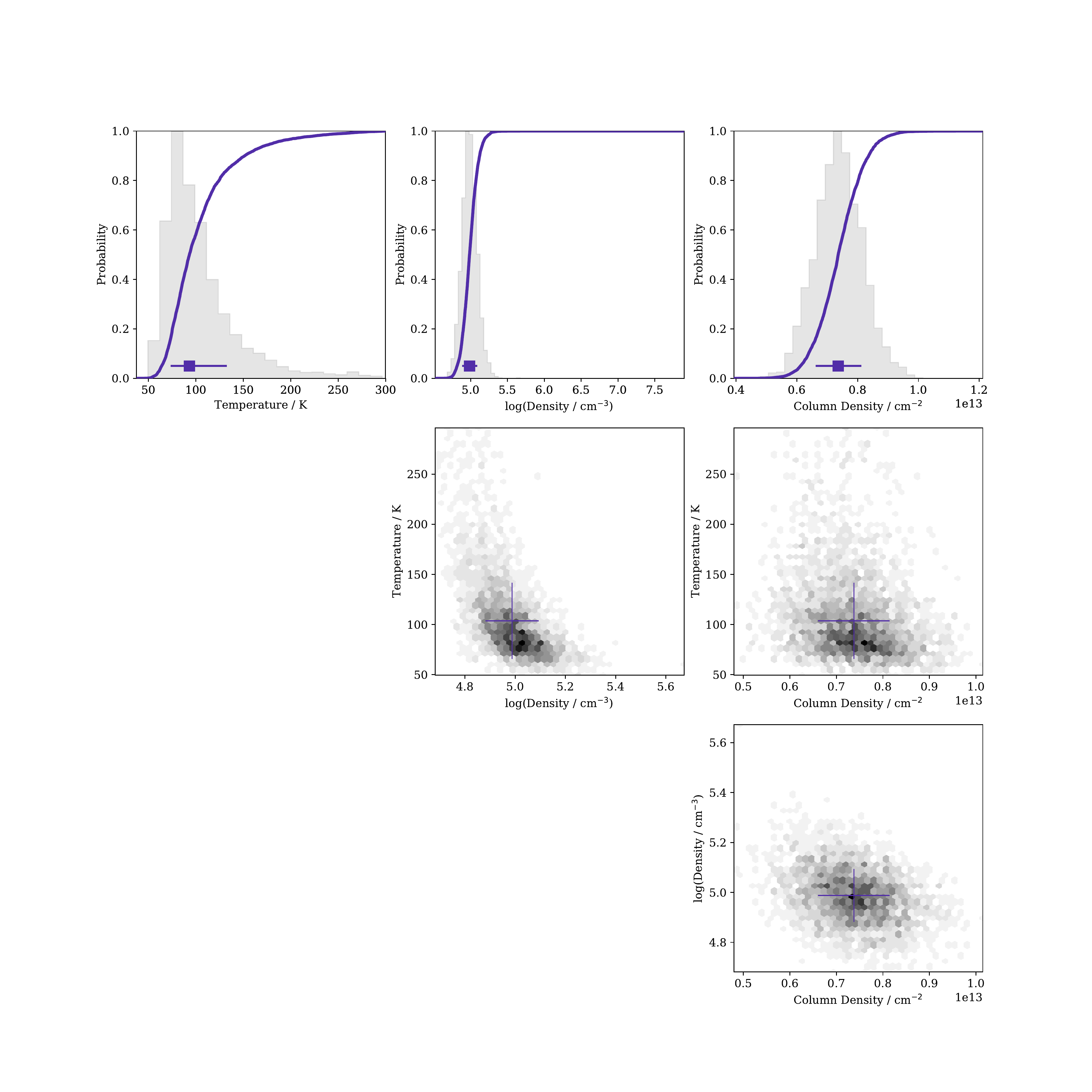}
\caption{As Figure~\ref{fig:h2cschain} but for p-H$_2$CS. The peak in the marginalized probability distribution for the gas density is consistent with the main peak in the corresponding o-H$_2$CS distribution. However, the density is much better constrained and the temperature-density degeneracy has been broken.\label{fig:ph2cschain}}
\end{figure*}

\begin{figure*}
\includegraphics[width=1.0\textwidth]{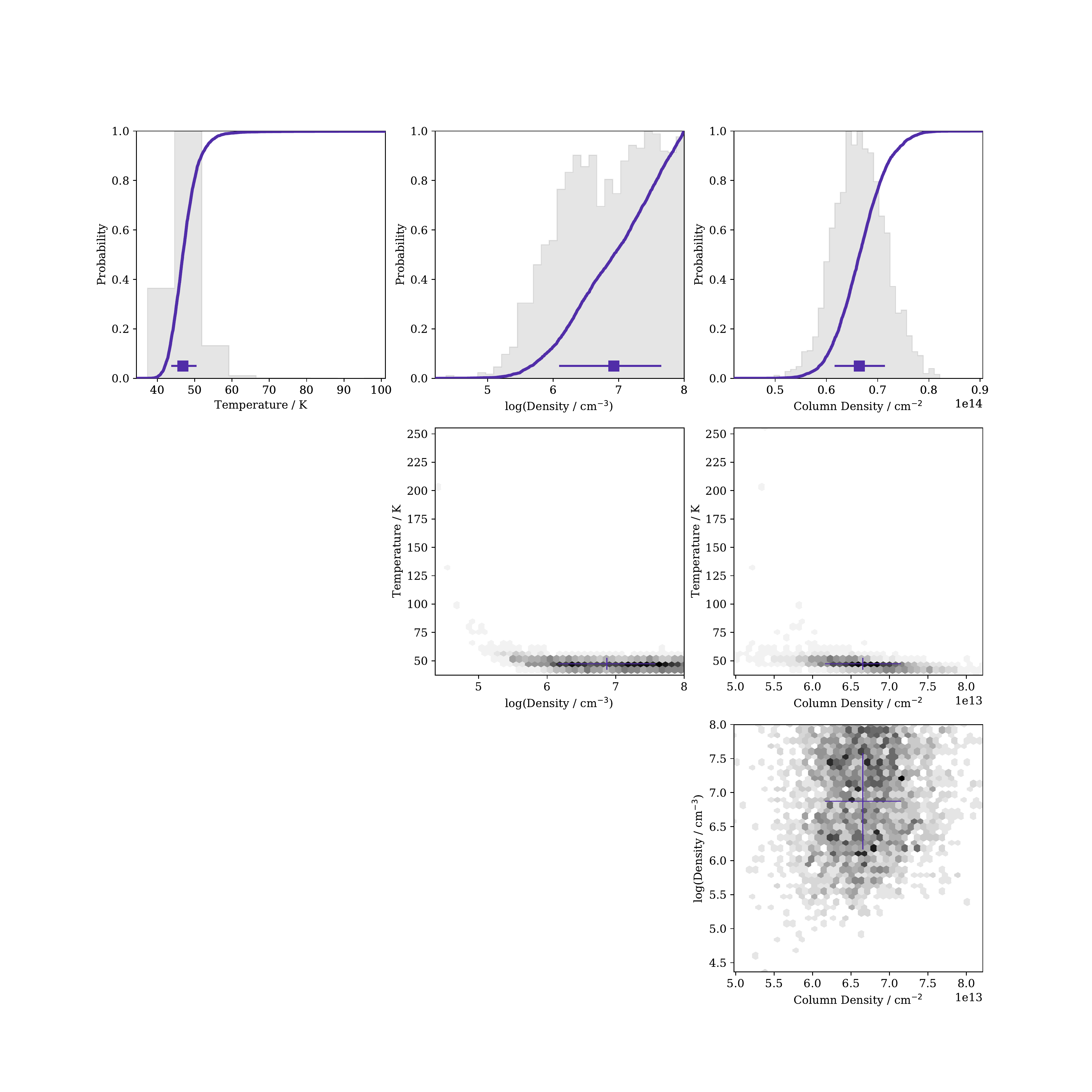}
\label{fig:ocschain}
\caption{As Figure~\ref{fig:h2cschain} but for OCS. Only a lower limit is found for the gas density but there is a strong peak at low temperatures. However, once again the probability distribution for the column density has a clear peak that is not dependent on the gas density.}
\end{figure*}

\begin{figure*}
\includegraphics[width=1.0\textwidth]{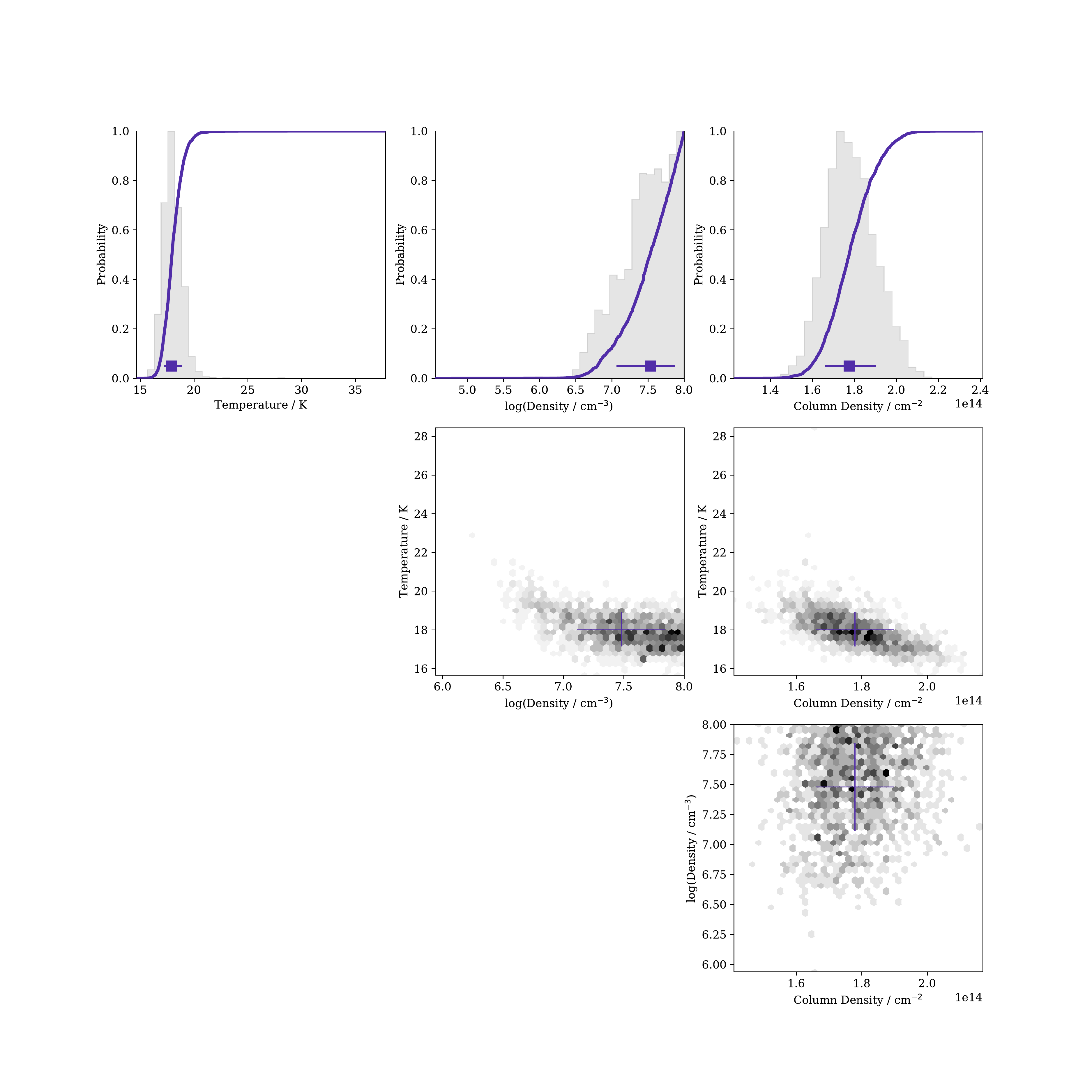}
\label{fig:sochain}
\caption{As Figure~\ref{fig:h2cschain} but for SO. Similar to OCS, SO shows only a lower limit on the density. Nevertheless, the SO column density has a clear most likely value. It also gives a strong peak at temperatures much lower than other molecules.}
\end{figure*}

\begin{figure*}
\includegraphics[width=1.0\textwidth]{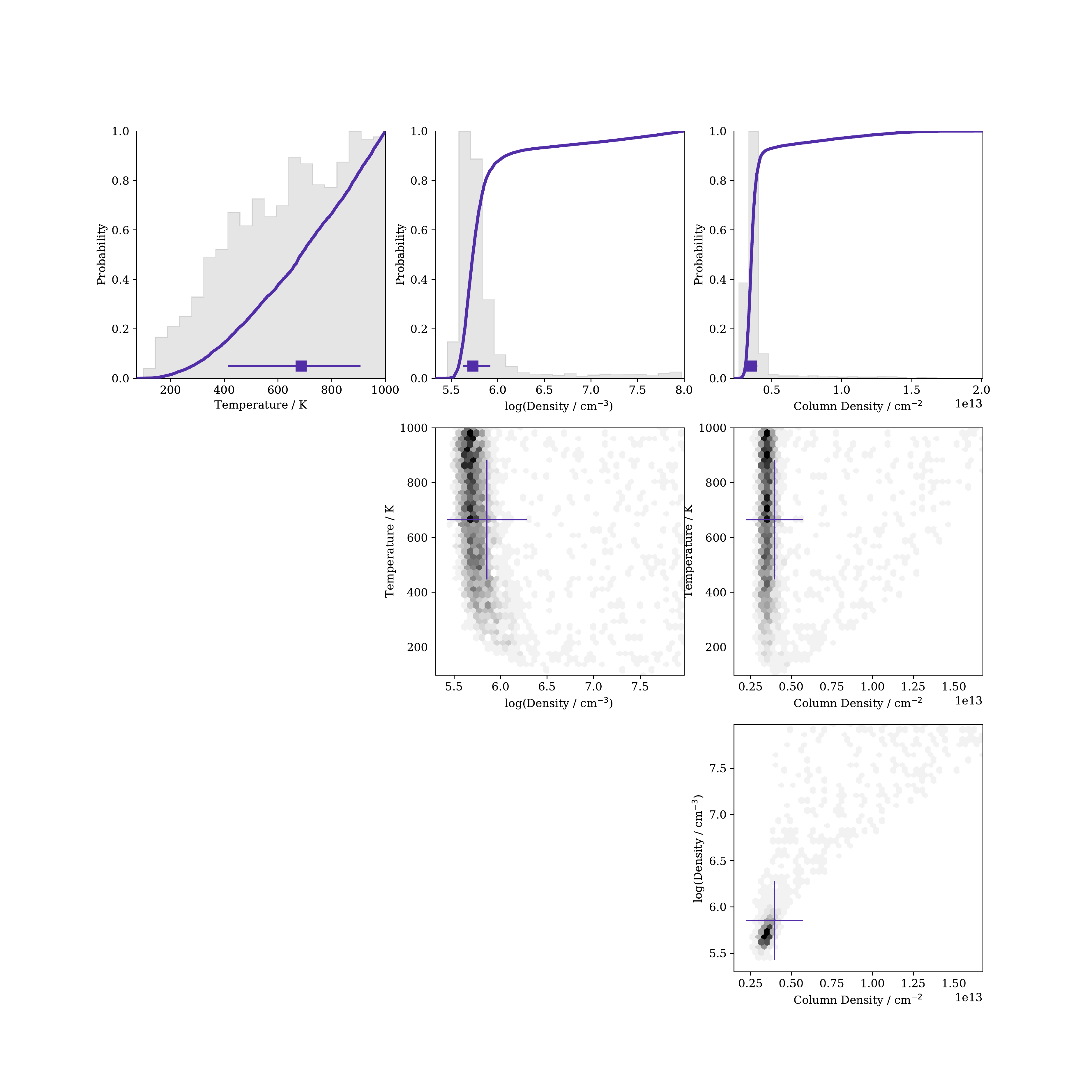}
\label{fig:sobumpchain}
\caption{As Figure~\ref{fig:h2cschain} but for high velocity SO emission. All parameters are well constrained though density and temperature appear to be strongly correlated. The central panel shows the joint probability distribution of the gas temperature and density, it is similar to the $\chi^2$ distribution for the secondary emission of CS in B1a shown in Figure 7 of \citet{Benedettini2013}, possibly indicating the secondary SO emission comes from the same region (See Section~\ref{sec:sosecondary}.}
\end{figure*}

\begin{figure*}
\includegraphics[width=1.0\textwidth]{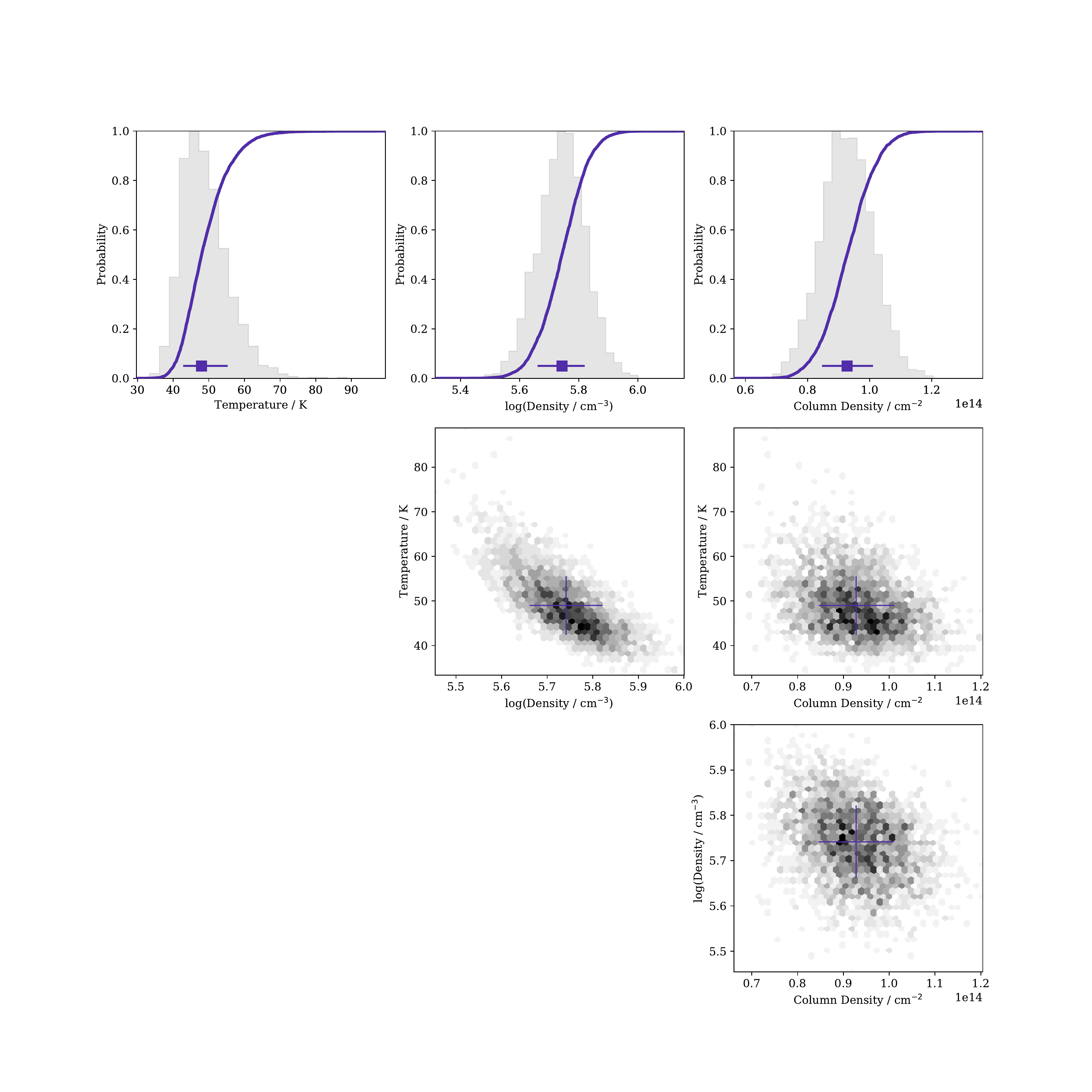}
\caption{As Figure~\ref{fig:h2cschain} but for SO$_2$. SO$_2$ is unique in that it shows a single peaked distribution for all input parameters.\label{fig:so2chain}}
\end{figure*}

Figures~\ref{fig:h2cschain} to \ref{fig:so2chain} show the marginalized probability distributions for the temperature, gas density and column density for H$_2$CS, OCS, SO and SO$_2$. These are given as histograms overplotted with a line representing the cumulative probability distribution. The marginalized probability distribution for each parameter shows the relative probability of any value given that we model the emission as a single gas component in RADEX. Error bars are used to show the median of the distributions and intervals containing 67\% of the total probability.\par
The joint probability distributions for each pair of variables are also plotted as colourmaps where darker areas correspond to higher likelihoods. These show any correlations between parameters. For example, in many cases high temperature values are more viable when combined with low gas densities and vice versa. A key result to note from these distributions is the lack of correlation between the column density and the other variables for each species. This indicates that the value derived for the column density is largely independent of the uncertainties in the gas properties.

\end{document}